\documentclass[11pt]{article}
\usepackage{graphicx}
\usepackage{amsmath,amssymb,amsfonts,amsthm}
\usepackage{setspace}
\usepackage{subfig}
\usepackage{booktabs}
\usepackage{multirow}
\usepackage{hyperref}
\usepackage{float}

\doublespace

\numberwithin{equation}{section}

\usepackage[top=1.25 in, bottom=1.25 in, left=1.25 in , right=1.25 in]{geometry}

\newtheorem{theorem}{Theorem}
\newtheorem{lemma}{Lemma}

\newtheorem{assumption}{Assumption}

\newenvironment{example}[1][Example]{\begin{trivlist}
\item[\hskip \labelsep {\bfseries #1}]}{\end{trivlist}}

\DeclareMathOperator*{\argmin}{arg\,min}

\makeatletter
\renewcommand\@biblabel[1]{}
\makeatother

\hypersetup{
    colorlinks,
    citecolor=black,%
    filecolor=black,%
    linkcolor=black,%
    urlcolor=blue}

\begin{document}

\title{\Large{Asymptotic Refinements of a Misspecification-Robust Bootstrap for Generalized Method of Moments Estimators}}

\author{Seojeong Lee
\footnote{School of Economics, Australian School of Business, University of New South Wales, Sydney, NSW 2052 Australia, Tel.: (+61) 2 9385 3325, Fax: (+61) 2 9313 6337, Email: \href{mailto:jay.lee@unsw.edu.au}{jay.lee@unsw.edu.au}, Homepage: \href{sites.google.com/site/misspecifiedjay/}{sites.google.com/site/misspecifiedjay/}}\\
\textit{\normalsize{University of New South Wales}}}

\date{\normalsize{Accepted for publication at the Journal of Econometrics}}

\maketitle

\vspace{-1em}

\begin{abstract}
I propose a nonparametric iid bootstrap that achieves asymptotic refinements for $t$ tests and confidence intervals based on GMM estimators even when the model is misspecified. In addition, my bootstrap does not require recentering the moment function, which has been considered as critical for GMM. Regardless of model misspecification, the proposed bootstrap achieves the same sharp magnitude of refinements as the conventional bootstrap methods which establish asymptotic refinements by recentering in the absence of misspecification.  The key idea is to link the misspecified bootstrap moment condition to the large sample theory of GMM under misspecification of Hall and Inoue (2003). Two examples are provided: Combining data sets and invalid instrumental variables.\\

\end{abstract}

%%%%%%%%%%%%%%%%%%%%%%%%%%%%%%%%%%%%%%%%%%%%%%%%%%%%%%%%%
\section{Introduction}
%%%%%%%%%%%%%%%%%%%%%%%%%%%%%%%%%%%%%%%%%%%%%%%%%%%%%%%%%

This paper proposes a novel bootstrap procedure for the generalized method of moments (GMM) estimators of Hansen (1982). It extends the existing literature by establishing the same asymptotic refinements for $t$ tests and confidence intervals (CI's) (i) without recentering the bootstrap moment function, and (ii) without assuming correct model specification. In contrast, the conventional bootstrap achieves the refinements only if recentering is done and the assumed moment condition is correctly specified. Thus, the contribution of this paper may look too good to be true at first glance, but it becomes apparent once we realize that those two eliminations are in fact closely related, because recentering makes the bootstrap non-robust to misspecification.

Bootstrapping has been considered as an alternative to the first-order GMM asymptotic theory, which has been known to provide poor approximations of finite sample distributions of test statistics especially when the model is highly non-linear or the number of moments is large, e.g., Blundell and Bond (1998), Bond and Windmeijer (2005), Hansen, Heaton, and Yaron (1996), Kocherlakota (1990), and Tauchen (1986).\footnote{The 1996 special issue of the \textit{Journal of Business \& Economic Statistics} deals with this problem in various contexts.} Hahn (1996) proves the first-order validity of the bootstrap distribution of GMM estimators. Hall and Horowitz (1996) show asymptotic refinements of the bootstrap for $t$ tests and the $J$ test (henceforth the Hall-Horowitz bootstrap). Andrews (2002) proposes a computationally attractive $k$-step bootstrap procedure based on the Hall-Horowitz bootstrap. Inoue and Shintani (2006) extend the Hall-Horowitz bootstrap by allowing correlation of moment functions beyond finitely many lags. Brown and Newey (2002) suggest an alternative bootstrap procedure using the empirical likelihood (EL) probability (henceforth the Brown-Newey bootstrap).

In the existing bootstrap methods for GMM estimators, recentering is critical. Horowitz (2001) explains why recentering is important when applying the bootstrap to overidentified moment condition models, where the dimension of a moment function is greater than that of a parameter. In such models, the sample mean of the moment function evaluated at the estimator is not necessarily equal to zero, though it converges almost surely to zero if the model is correctly specified. In principle, the bootstrap considers the sample and the estimator as if they were the population and the true parameter, respectively. This implies that the bootstrap version of the moment condition, that the sample mean of the moment function evaluated at the estimator should equal to zero, does not hold when the model is overidentified. Recentering makes the bootstrap version of the moment condition hold. The Hall-Horowitz bootstrap analytically recenters the bootstrap moment function with respect to the sample moment condition. The Brown-Newey bootstrap recenters the bootstrap moment condition by employing the EL probability in resampling the bootstrap sample. Thus, both the Hall-Horowitz bootstrap and the Brown-Newey bootstrap can be referred as \textit{the recentered bootstrap}.

A naive bootstrap is to apply the standard bootstrap procedure as is done for just-identified models, without any additional correction, such as recentering. However, it turns out that this naive bootstrap fails to achieve asymptotic refinements for $t$ tests and CI's, and jeopardizes first-order validity of the $J$ test. Hall and Horowitz (1996) and Brown and Newey (2002) explain that the bootstrap and sample versions of test statistics would have different asymptotic distributions without recentering, because of the violation of the moment condition in the sample.

Although they address that the failure of the naive bootstrap is due to the misspecification in the sample, they do not further investigate the conditional asymptotic distribution of the bootstrap GMM estimator under misspecification. Instead, they eliminate the misspecification problem by recentering. In contrast, I observe that the conditional asymptotic covariance matrix of the bootstrap GMM estimator under misspecification is different from the standard one. The conditional asymptotic covariance matrix is consistently estimable by using the result of Hall and Inoue (2003), and I construct the $t$ statistic of which distribution is asymptotically standard normal even under misspecification.

Hall and Inoue (2003) show that the asymptotic distributions of GMM estimators under misspecification are different from those of the standard GMM theory.\footnote{Hall and Inoue (2003) does not deal with bootstrapping, however.} In particular, the asymptotic covariance matrix has additional non-zero terms in the presence of misspecification. Hall and Inoue's formulas for the asymptotic covariance matrix encompass the case of correct specification as a special case. The variance estimator using their formula is denoted by the Hall-Inoue variance estimator, hereinafter. Imbens (1997) also describes the asymptotic covariance matrices of GMM estimators robust to misspecification by using a just-identified formulation of overidentified GMM. However, his description is general, rather than being specific to the misspecification problem defined in this paper.

I propose a bootstrap procedure that uses the Hall-Inoue variance estimators in constructing the sample and the bootstrap $t$ statistics. It ensures that the bootstrap $t$ statistic satisfies the asymptotic pivotal condition without recentering. Moreover, the sample $t$ statistic is also asymptotically pivotal regardless of misspecification in the population. In other words, my bootstrap applies to the robust $t$ statistic which is studentized with the Hall-Inoue variance estimator. Therefore, it works without assuming correct model specification in the population, and is referred to as the misspecification-robust (MR) bootstrap. In contrast, the conventional first-order asymptotics as well as the recentered bootstrap would not work under misspecification, because the conventional $t$ statistic is not asymptotically pivotal anymore.

The MR bootstrap achieves asymptotic refinements, a reduction in the error of test rejection probability and CI coverage probability by a factor of $n^{-1}$ for symmetric two-sided $t$ tests and symmetric percentile-$t$ CI's, over the asymptotic counterparts. The magnitude of the error is $O(n^{-2})$, which is sharp. This is the same magnitude of error shown in Andrews (2002), that uses the Hall-Horowitz bootstrap for independent and identically distributed (iid) data with slightly stronger assumptions than those of Hall and Horowitz (1996).

I note that the MR bootstrap is not for the $J$ test. To get the bootstrap distribution of the $J$ statistic, the bootstrap should be implemented under the null hypothesis that the model is correctly specified. The recentered bootstrap imposes the null hypothesis of the $J$ test because it eliminates the misspecification in the bootstrap world by recentering. In contrast, the MR bootstrap does not eliminate the misspecification and thus, it does not mimic the distribution of the $J$ statistic under the null. Since the conventional asymptotic and bootstrap $t$ tests and CI's are valid only in the absence of misspecification, it is important to conduct the $J$ test and report the result that the model is not rejected. However, even a significant $J$ statistic would not invalidate the estimation results if possible misspecification of the model is assumed and the validity of $t$ tests and CI's is established under such assumption, as is done in this paper.

Three papers in the literature are in a similar vein in terms of bootstrap methods under misspecification. Corradi and Swanson (2006) show the first-order validity of the block bootstrap for conditional distribution tests under dynamic misspecification. Kline and Santos (2012) examine the higher-order properties of the wild bootstrap in a linear regression model when the mean independent assumption of the error term is misspecified. In particular, a referee suggested to clarify the marginal contribution of this paper with respect to the work of Gon\c{c}alves and White (2004) which proves the first-order validity of the bootstrap for $t$ tests based on the quasi-maximum likelihood (QML) estimators studentized with the misspecification-robust variance estimator of White (1982). 

First, the QML estimator is a special case of the GMM estimator when one uses the first-order condition of the QML as the moment condition. This also puts an additional restriction that the model is just-identified. Therefore, this paper covers a broader class of models than Gon\c calves and White (2004). For example, the proposed bootstrap applies to the two-stage least squares (2SLS) estimator. In addition, the definition of misspecified moment condition model should be distinguished from that of misspecified likelihood function. The former arises only when the model is overidentified, which implies that the first-order condition of the QML forms a correctly specified moment condition even if the likelihood function is misspecified. Thus, the misspecification-robust QML variance estimator corresponds to the conventional GMM variance estimator under correct specification, rather than the Hall-Inoue variance estimator.\footnote{Hall and Inoue (2003) explain their marginal contribution over Gallant and White (1988), White (1996), and Maasoumi and Phillips (1982) in this regard.}

Second, Gon\c calves and White (2004) neither provide a guidance whether to recenter or not, nor explain the relationship between recentering and misspecification. One of the contributions of Hall and Horowitz (1996) is that bootstrapping for GMM is non-standard so that one should recenter the moment function to achieve asymptotic refinements. I argue that recentering can be detrimental and is not even needed if we use the Hall-Inoue variance estimator. The key idea is to link the misspecified moment condition in the bootstrap world to the large sample theory of GMM under misspecification of Hall and Inoue (2003).

The remainder of the paper is organized as follows. Section \ref{Robust} discusses theoretical and empirical implications of misspecified models and explains the advantage of using the MR bootstrap $t$ tests and CI's. Section \ref{Outline} outlines the main result. Section \ref{Est} defines the estimators and test statistics. Section \ref{MRB} defines the nonparametric iid MR bootstrap for iid data. Section \ref{Main} states the assumptions and establishes asymptotic refinements of the MR bootstrap. Section \ref{MC} presents Monte Carlo simulation results. Section \ref{Conc} concludes the paper. Lemmas and proofs are gathered in the Appendix.

%%%%%%%%%%%%%%%%%%%%%%%%%%%%%%%%%%%%%%%%%%%%%%%%%%%%%%%%%%%
\section{Why We Care About Misspecification}
\label{Robust}
%%%%%%%%%%%%%%%%%%%%%%%%%%%%%%%%%%%%%%%%%%%%%%%%%%%%%%%%%%%

Empirical studies in the economics literature often report a significant $J$ statistic along with GMM estimates, standard errors, and CI's. Such examples include Imbens and Lancaster (1994), Jondeau, Le Bihan, and Galles (2004), Parker and Julliard (2005), and Ag\"{u}ero and Marks (2008), among others. Significant $J$ statistics are also quite common in the instrumental variables literature using the 2SLS estimator, which is a special case of the GMM estimator.

A significant $J$ statistic means that the test rejects the null hypothesis of correct model specification. For 2SLS estimators, this implies that at least one of the instruments is invalid. The problem is that, even if models are likely to be misspecified, inferences are made using the asymptotic theory for correctly specified models and the estimates are interpreted with economic implications. Various authors justify this by noting that the $J$ test over-rejects the correct null in small samples.

On the other hand, comparing and evaluating the relative fit of competing models have been an important research topic. Vuong (1989), Rivers and Vuong (2002), and Kitamura (2003) suggest various tests of the null hypothesis that test whether two possibly misspecified models provide equivalent approximation to the true model in terms of the Kullback-Leibler information criteria (KLIC). Recent studies such as Chen, Hong, and Shum (2007), Marmer and Otsu (2012), and Shi (2013) generalize and modify the test in broader settings. Hall and Pelletier (2011) show that the limiting distribution of the Rivers-Vuong test statistic may not be consistently estimable unless both models are misspecified. In this framework, therefore, all competing models are misspecified and the test selects a less misspecified model. For applications of the Rivers-Vuong test, see French and Jones (2004), Gowrisankaran and Rysman (2009), and Bonnet and Dubois (2010).

Either for the empirical studies that report a significant $J$ statistic, or for a model selected by the Rivers-Vuong test, inferences about the parameters should take into account a possible misspecification in the model. Otherwise, such inferences would be misleading.

\begin{example}[Example: Combining Micro and Macro Data]\

Imbens and Lancaster (1994) suggest an econometric procedure that uses nearly exact information on the marginal distribution of economic variables to improve accuracy of estimation. As an application, the authors estimate the following probit model for employment: For an individual $i$,
\begin{eqnarray}
P(L_{i}=1|Age_{i},Edu_{i})&=&\Phi(X_{i}'\theta)\\
\nonumber &=&\Phi(\theta_{0}+\theta_{1}\cdot Edu_{i} + \theta_{2}\cdot(Age_{i}-35) + \theta_{3}\cdot (Age_{i}-35)^{2}),
\end{eqnarray}
with $X_{i}=(1,Edu_{i},Age_{i}-35,(Age_{i}-35)^{2})'$ and $\Phi(\cdot)$ is the standard normal cdf. $L_{i}$ is labor market status ($L_{i}=1$ when employed), $Edu_{i}$ is education level in five categories, and $Age_{i}$ is age in years. The sample is a micro data set on Dutch labor market histories and the number of observations is 347. Typically, the probit model is estimated by the ML estimator. The first row of Table \ref{IL1994} presents the ML point estimates and the standard errors. None of the coefficients are statistically significant except for that of the intercept.

To reduce the standard errors of the estimators, the authors use additional information on the population from the national statistics. By using the statistical yearbooks for the Netherlands which contain 2.355 million observations, they calculated the probability of being employed given the age category (denoted by $p_{k}$ where the index for the age category $k=1,2,3,4,5$) and the probability of being in a particular age category (denoted by $q_{k}$). These probabilities are considered as the true population parameters.

The authors suggest to use GMM estimators with the moment function that utilizes the information from the aggregate statistic. The second row of Table \ref{IL1994} reports the two-step efficient GMM point estimates and the standard errors. Now the coefficient $\theta_{3}$ is statistically significant at 1\% level and the authors argue ``...Age is not ancillary anymore and knowledge about its marginal distribution is informative about $\theta$.''

Although they could successfully improve the accuracy of the estimators by combining two data sets, their argument has a potential problem. The last column of Table \ref{IL1994} reports the $J$ test statistic and its $p$-value. Since the $p$-value is 4.4\%, the model is marginally rejected at 5\% level. The problem is that, if the model is truly misspecified, the reported GMM standard errors are inconsistent because the conventional standard errors are only consistent under correct specification. Then the authors' argument about the coefficient estimates may be flawed. This problem could be avoided if the standard errors which are consistent even under misspecification were used. The formulas for the misspecification-robust standard errors for the GMM estimators are available in Section 4.\footnote{Since the original data sets used in Imbens and Lancaster (1994) are not available, I could not calculate the robust standard errors. Instead, I provide simulation result with a simple hypothetical model that utilizes additional population information in estimation in Section 7.1.}
\end{example}

When the model is misspecified, $Eg(X_{i},\theta)\neq 0$ for all $\theta$, where $\theta$ is a parameter of interest, $X_{i}$ is a random vector, $g(X_{i},\theta)$ is a known moment function, and $E[\cdot]$ denotes mathematical expectation. Let $\hat{\theta}$ be the GMM estimator and $\Omega^{-1}$ be a positive definite matrix, which is the probability limit of a weight matrix. According to Hall and Inoue (2003), (i) the probability limit of $\hat{\theta}$ is the pseudo-true value that depends on $\Omega^{-1}$ such that
\begin{equation}
\theta_{0}(\Omega^{-1}) = \argmin_{\theta} Eg(X_{i},\theta)'\Omega^{-1}Eg(X_{i},\theta),
\label{Pcrit}
\end{equation}
and (ii) the asymptotic distribution of the GMM estimator is
\begin{equation}
\sqrt{n}(\hat{\theta}-\theta_{0}(\Omega^{-1}))\rightarrow_{d}N(0,\Sigma_{MR}),
\label{secondC}
\end{equation}
where $\Sigma_{MR}$ is the asymptotic covariance matrix under misspecification that is different from $\Sigma_{C}$, the asymptotic covariance matrix under correct specification. If the model is correctly specified, then $\theta_{0}(\Omega^{-1})$ and $\Sigma_{MR}$ simplify to $\theta_{0}$ and $\Sigma_{C}$, respectively.

The pseudo-true value can be interpreted as the best approximation to the true value, if any, given the weight matrix. The dependence of the pseudo-true value on the weight matrix may make the interpretation of the estimand unclear. Nevertheless, the literature on estimation under misspecification considers the pseudo-true value as a valid estimand, see Sawa (1978), White (1982), and Schennach (2007) for more discussions. Other pseudo-true values that minimize the generalized empirical likelihood (GEL) without using a weight matrix, have better interpretations but comparing different pseudo-true values is beyond the scope of this paper.

Although we cannot fix a potential bias in the pseudo-true value in general, we can report the standard error of the GMM estimator as honest as possible. \eqref{secondC} implies that the conventional $t$ tests and CI's are invalid under misspecification, because the conventional standard errors are based on the estimate of $\Sigma_{C}$. Misspecification-robust standard errors are calculated using the Hall-Inoue variance estimator of $\Sigma_{MR}$. By using the robust standard errors, the resulting asymptotic $t$ tests and CI's are robust to misspecification. The MR bootstrap $t$ tests and CI's improve upon these MR asymptotic $t$ tests and CI's in terms of the magnitude of errors in test rejection probability and CI coverage probability. A summary on the advantage of the MR bootstrap over the existing asymptotic and bootstrap $t$ tests and CI's is given in Table \ref{COMP}.

One may consider local misspecification to model a slight misspecification which may not be detected by the $J$ test. A recent development on this topic includes the works of Bravo (2010), Berkowitz, Caner, and Fang (2008, 2012), DiTraglia (2012), Guggenberger (2012), Guggenberger and Kumar (2012), Hall (2005), and Otsu (2011). Local misspecification enables us to make a better interpretation of the pseudo-true value. To see this, let a triangular array $\{X_{n,i}\}_{i\leq n}$ be iid over $i$ for fixed $n$, where $n$ is the sample size. The moment condition is locally misspecified if
\begin{equation}
\nonumber Eg(X_{n,i},\theta_{0}) = \frac{\delta}{\sqrt{n}},
\end{equation}
where $\theta_{0}$ is a true parameter and $\delta$ is an unknown vector of constants. Since the GMM estimator $\hat{\theta}$ is not $\sqrt{n}$-consistent for $\theta_{0}$ in this setting, the MR bootstrap CI as well as the conventional CI's does not give asymptotically correct coverage for $\theta_{0}$. 

%%%%%%%%%%%%%%%%%%%%%%%%%%%%%%%%%%%%%%%%%%%%%%%%%%
\section{Outline of the Results}
\label{Outline}
%%%%%%%%%%%%%%%%%%%%%%%%%%%%%%%%%%%%%%%%%%%%%%%%%%

In this section, I outline the MR bootstrap. The idea of the MR bootstrap procedure can be best understood in the same framework with Hall and Horowitz (1996) and Brown and Newey (2002), as is described below.

Suppose that the random sample is $\chi_{n}=\{X_{i}:i\leq n\}$ from a probability distribution $P$. Let $F$ be the corresponding cumulative distribution function (cdf). The empirical distribution function (edf) is denoted by $F_{n}$. The GMM estimator, $\hat{\theta}$, minimizes a sample criterion function, $J_{n}(\theta)$. Suppose that $\theta$ is a scalar for notational brevity. Let $\hat{\Sigma}$ be a consistent estimator of the asymptotic variance of $\sqrt{n}(\hat{\theta}-plim(\hat{\theta}))$.

I also define the bootstrap sample. Let $\chi_{n_{b}}^{*}=\{X_{i}^{*}:i\leq n_{b}\}$ be a sample of random vectors from the empirical distribution $P^{*}$ conditional on $\chi_{n}$ with the edf $F_{n}$. In this section, I distinguish $n$ and $n_{b}$, which helps to understand the concept of the conditional asymptotic distribution.\footnote{$n_{b}$ is the resample size and should be distinguished from the number of bootstrap replication (or resampling), often denoted by $B$. See Bickel and Freedman (1981) for further discussion.} I set $n=n_{b}$ from the following section. Define $J_{n_{b}}^{*}(\theta)$ and $\hat{\Sigma}^{*}$ like $J_{n}(\theta)$ and $\hat{\Sigma}$ are defined, but with $\chi_{n_{b}}^{*}$ in place of $\chi_{n}$. The bootstrap GMM estimator $\hat{\theta}^{*}$ minimizes $J_{n_{b}}^{*}(\theta)$.

Consider a symmetric two-sided test of the null hypothesis $H_{0}:\theta=\theta_{0}$ with level $\alpha$. The $t$ statistic under $H_{0}$ is $T(\chi_{n})=(\hat{\theta}-\theta_{0})/\sqrt{\hat{\Sigma}/n}$, a functional of $\chi_{n}$. One rejects the null hypothesis if $|T(\chi_{n})|>z$ for a critical value $z$. I also consider a $100(1-\alpha)\%$ CI for $\theta_{0}$, $[\hat{\theta}\pm z \sqrt{\hat{\Sigma}/n}]$. For the asymptotic test or the asymptotic CI, set $z=z_{\alpha/2}$, where $z_{\alpha/2}$ is the $1-\alpha/2$ quantile of a standard normal distribution. For the bootstrap test or the symmetric percentile-$t$ interval, set $z=z^{*}_{|T|,\alpha}$, where $z^{*}_{|T|,\alpha}$ is the $1-\alpha$ quantile of the distribution of $|T(\chi^{*}_{n_{b}})|\equiv |\hat{\theta}^{*}-\hat{\theta}|/\sqrt{\hat{\Sigma}^{*}/n_{b}}$.

Let $H_{n}(z,F)=P(T(\chi_{n})\leq z|F)$ and $H_{n_{b}}^{*}(z,F_{n})=P(T(\chi_{n_{b}}^{*})\leq z|F_{n})$. According to Hall (1992), under regularity conditions, $H_{n}(z,F)$ and $H_{n_{b}}^{*}(z,F_{n})$ allow Edgeworth expansion of the form
\begin{eqnarray}
H_{n}(z,F) &=& H_{\infty}(z,F) + n^{-1/2}q_{1}(z,F) + n^{-1}q_{2}(z,F) + o(n^{-1}),
\label{Ed1}\\
H_{n_{b}}^{*}(z,F_{n}) &=& H_{\infty}^{*}(z,F_{n}) + n_{b}^{-1/2}q_{1}(z,F_{n}) + n_{b}^{-1}q_{2}(z,F_{n}) + o_{p}(n_{b}^{-1})
\label{Ed2}
\end{eqnarray}
uniformly over $z$, where $q_{1}(z,F)$ is an even function of $z$ for each $F$, $q_{2}(z,F)$ is an odd function of $z$ for each $F$, $q_{2}(z,F_{n})\rightarrow q_{2}(z,F)$ almost surely as $n\rightarrow\infty$ uniformly over $z$, $H_{\infty}(z,F)=\lim_{n\rightarrow\infty}H_{n}(z,F)$ and $H_{\infty}^{*}(z,F_{n})=\lim_{n_{b}\rightarrow\infty}H_{n_{b}}^{*}(z,F_{n})$. If $T(\cdot)$ is asymptotically pivotal, then $H_{\infty}(z,F)=H_{\infty}^{*}(z,F_{n})=\Phi(z)$ where $\Phi$ is the standard normal cdf, because $H_{\infty}(z,F)$ and $H_{\infty}^{*}(z,F_{n})$ do not depend on the underlying cdf.

Using \eqref{Ed1} and the fact that $q_{1}$ is even, it can be shown that under $H_{0}$,
\begin{equation}
P(|T(\chi_{n})|>z_{\alpha/2})=\alpha+O(n^{-1}),\hspace{1em}P(\theta_{0}\in CI)=1-\alpha+O(n^{-1}),
\label{ASrate}
\end{equation}
where $CI=[\hat{\theta}\pm z_{\alpha/2}\sqrt{\hat{\Sigma}/n}]$. In other words, the error in the rejection probability and coverage probability of the asymptotic two-sided $t$ test and CI is $O(n^{-1})$.

For the bootstrap $t$ test and CI, subtract \eqref{Ed1} from \eqref{Ed2}, use the fact that $q_{1}$ is even, and set $n_{b}=n$ to show, under $H_{0}$,
\begin{equation}
P(|T(\chi_{n})|>z^{*}_{|T|,\alpha})=\alpha+o(n^{-1}),\hspace{1em}P(\theta_{0}\in CI^{*})=1-\alpha+o(n^{-1})
\label{BTrate0}
\end{equation}
where $CI^{*}=[\hat{\theta} \pm z^{*}_{|T|,\alpha}\sqrt{\hat{\Sigma}/n}]$. The elimination of the leading terms in \eqref{Ed1} and \eqref{Ed2} is the source of asymptotic refinements of bootstrapping the asymptotically pivotal statistics (Beran, 1988; Hall, 1992).

First, suppose that the model is correctly specified, $Eg(X_{i},\theta_{0})=0$ for unique $\theta_{0}$, where $E[\cdot]$ is the expectation with respect to the cdf F. The conventional $t$ statistic $T_{C}(\chi_{n})=(\hat{\theta}-\theta_{0})/\sqrt{\hat{\Sigma}_{C}/n}$, where $\hat{\Sigma}_{C}$ is the standard GMM variance estimator, is asymptotically pivotal. However, a naive bootstrap $t$ statistic without recentering,\footnote{A naive bootstrap for GMM is constructing $\hat{\theta}^{*}$ and $\hat{\Sigma}^{*}$ in the same way we construct $\hat{\theta}$ and $\hat{\Sigma}$, using the bootstrap sample $\chi_{n_{b}}^{*}$ in place of $\chi_{n}$.} $T_{C}(\chi_{n_{b}}^{*})=(\hat{\theta}^{*}-\hat{\theta})/\sqrt{\hat{\Sigma}_{C}^{*}/n_{b}}$, is not asymptotically pivotal because the moment condition under $F_{n}$ is misspecified, $E_{F_{n}}g(X_{i}^{*},\hat{\theta})=n^{-1}\sum_{i=1}^{n}g(X_{i},\hat{\theta})\neq 0$ almost surely when the model is overidentified, where $E_{F_{n}}[\cdot]$ is the expectation with respect to $F_{n}$. If the moment condition is misspecified, the conventional GMM variance estimator is no longer consistent. Note that the bootstrap moment condition is evaluated at $\hat{\theta}$, where $\hat{\theta}$ is considered as the true value given $F_{n}$.

The recentered bootstrap makes the bootstrap moment condition hold so that the recentered bootstrap $t$ statistic is asymptotically pivotal. For instance, the Hall-Horowitz bootstrap uses a recentered moment function $g^{*}(X_{i}^{*},\theta)=g(X_{i}^{*},\theta)-n^{-1}\sum_{i=1}^{n}g(X_{i},\hat{\theta})$ so that $E_{F_{n}}g^{*}(X_{i}^{*},\hat{\theta})=0$ almost surely. The Brown-Newey bootstrap uses the EL distribution function $\hat{F}_{EL}(z)=n^{-1}\sum_{i=1}^{n}\hat{p}_{i}\mathbf{1}(X_{i}\leq z)$ in resampling, where $\hat{p}_{i}$ is the EL probability and $\mathbf{1}(\cdot)$ is an indicator function, instead of using $F_{n}$, so that $E_{\hat{F}_{EL}}g(X_{i}^{*},\hat{\theta})=0$ almost surely, where $E_{\hat{F}_{EL}}[\cdot]$ is the expectation with respect to $\hat{F}_{EL}$.

The MR bootstrap uses the original \textit{non-recentered} moment function in implementing the bootstrap and resamples according to the edf $F_{n}$. This is similar to the naive bootstrap. The distinction is that the MR bootstrap uses the Hall-Inoue variance estimator in constructing the sample and the bootstrap versions of the $t$ statistic instead of using the conventional GMM variance estimator. The sample $t$ statistic is $T_{MR}(\chi_{n})=(\hat{\theta}-\theta_{0})/\sqrt{\hat{\Sigma}_{MR}/n}$, where $\hat{\Sigma}_{MR}$ is a consistent estimator of $\Sigma_{MR}$, the asymptotic variance of the GMM estimator regardless of misspecification. $T_{MR}(\chi_{n})$ is asymptotically pivotal.

The MR bootstrap $t$ statistic is $T_{MR}(\chi_{n_{b}}^{*})=(\hat{\theta}^{*}-\hat{\theta})/\sqrt{\hat{\Sigma}_{MR}^{*}/n_{b}}$, where $\hat{\Sigma}_{MR}^{*}$ uses the same formula as $\hat{\Sigma}_{MR}$ with $\chi_{n_{b}}^{*}$ in place of $\chi_{n}$. $\hat{\Sigma}_{MR}^{*}$ is consistent for the conditional asymptotic variance of the bootstrap GMM estimator, $\Sigma_{MR|F_{n}}$, almost surely, even if the bootstrap moment condition is not satisfied. As a result, $T_{MR}(\chi_{n_{b}}^{*})$ is asymptotically pivotal. Therefore, the MR bootstrap achieves asymptotic refinements without recentering under correct specification.

Now suppose that the model is misspecified in the population, $Eg(X_{i},\theta)\neq 0$ for all $\theta$. The advantage of the MR bootstrap is that neither the sample $t$ statistic nor the bootstrap $t$ statistic requires the assumption of correct model. Since $T_{MR}(\chi_{n})$ and $T_{MR}(\chi_{n_{b}}^{*})$ are constructed by using the Hall-Inoue variance estimator, they are asymptotically pivotal regardless of model misspecification. Thus, the ability of achieving asymptotic refinements of the MR bootstrap is not affected.

The conclusion changes dramatically for the recentered bootstrap, however. First of all, the conventional $t$ statistic $T_{C}(\chi_{n})$ is no longer asymptotically pivotal and this invalidates the use of the asymptotic $t$ test and CI. Moreover, the recentered bootstrap $t$ test and CI are not first-order valid because (i) they use the inconsistent conventional standard error and, (ii) they impose a wrong moment condition by recentering.\footnote{The conditional and unconditional distributions of the recentered bootstrap $t$ statistic is described in Supplementary Appendix available at the author's webpage.}

Let $z^{*}_{|T_{MR}|,\alpha}$ be the $1-\alpha$ quantile of the distribution of $|T_{MR}(\chi^{*}_{n_{b}})|$ and let $CI_{MR}^{*}=[\hat{\theta} \pm z^{*}_{|T_{MR}|,\alpha}\sqrt{\hat{\Sigma}_{MR}/n}]$. Using the MR bootstrap without assuming the correct model, I show that, under $H_{0}$,
\begin{equation}
P(|T_{MR}(\chi_{n})|>z^{*}_{|T_{MR}|,\alpha})=\alpha+O(n^{-2}),\hspace{1em}P(\theta_{0}\in CI_{MR}^{*})=1-\alpha+O(n^{-2}).
\label{BTrate1}
\end{equation}
This rate is sharp. The further reduction in the error from $o(n^{-1})$ of \eqref{BTrate0} to $O(n^{-2})$ of \eqref{BTrate1} is based on the argument given in Hall (1988). Andrews (2002) shows the same sharp bound using the Hall-Horowitz bootstrap and assuming the correct model.

%%%%%%%%%%%%%%%%%%%%%%%%%%%%%%%%%%%%%%%%%%%%%%%%%%%%%%%%%%
\section{Estimators and Test Statistics}
\label{Est}
%%%%%%%%%%%%%%%%%%%%%%%%%%%%%%%%%%%%%%%%%%%%%%%%%%%%%%%%%%

Given an $L_{g}\times 1$ vector of moment conditions $g(X_{i},\theta)$, where $\theta$ is $L_{\theta}\times 1$, and $L_{g}\geq L_{\theta}$, define a correctly specified and a misspecified model as follows: The model is \textit{correctly specified} if there exists a unique value $\theta_{0}$ in $\Theta\subset\mathbb{R}^{L_{\theta}}$ such that $Eg(X_{i},\theta_{0})=0$, and the model is \textit{misspecified} if there exists no $\theta$ in $\Theta\subset\mathbb{R}^{L_{\theta}}$ such that $Eg(X_{i},\theta)=0$. That is, $Eg(X_{i},\theta)=g(\theta)$ where $g:\Theta\rightarrow\mathbb{R}^{L_{g}}$ such that $\|g(\theta)\|>0$ for all $\theta\in\Theta$, if the model is misspecified. Assume that the model is possibly misspecified.

The (pseudo-)true parameter $\theta_{0}$ minimizes the population criterion function,
\begin{equation}
J(\theta,\Omega^{-1})=Eg(X_{i},\theta)'\Omega^{-1} Eg(X_{i},\theta),
\end{equation}
where $\Omega^{-1}$ is the probability limit of a weight matrix. Since the model is possibly misspecified, the moment condition and the population criterion may not equal to zero for any $\theta\in\Theta$. In this case, the minimizer of the population criterion depends on $\Omega^{-1}$ and is denoted by $\theta_{0}(\Omega^{-1})$. We call $\theta_{0}(\Omega^{-1})$ the pseudo-true value. The dependence vanishes when the model is correctly specified.

Consider two forms of GMM estimator. The first one is a one-step GMM estimator using the identity matrix $I_{L_{g}}$ as a weight matrix, which is the common usage. The second one is a two-step GMM estimator using a weight matrix constructed from the one-step GMM estimator. Under correct specifications, the common choice of the weight matrix is an asymptotically optimal one. However, the optimality is not established under misspecification because the asymptotic covariance matrix of the two-step GMM estimator cannot be simplified to the efficient one under correct specification.

The one-step GMM estimator, $\hat{\theta}_{(1)}$, solves
\begin{equation}
\min_{\theta\in\Theta}J_{n}(\theta,I_{L_{g}})= \left(n^{-1}\sum_{i=1}^{n}g(X_{i},\theta)\right)'\left(n^{-1}\sum_{i=1}^{n}g(X_{i},\theta)\right).
\label{1gmm}
\end{equation}
The two-step GMM estimator, $\hat{\theta}_{(2)}$ solves
\begin{equation}
\min_{\theta\in\Theta}J_{n}(\theta,W_{n}(\hat{\theta}_{(1)}))\equiv \left(n^{-1}\sum_{i=1}^{n}g(X_{i},\theta)\right)'W_{n}(\hat{\theta}_{(1)})\left(n^{-1}\sum_{i=1}^{n}g(X_{i},\theta)\right),
\label{2gmm}
\end{equation}
where\footnote{One may consider an $L_{g}\times L_{g}$ nonrandom positive-definite symmetric matrix for the one-step GMM estimator or the \textit{uncentered} weight matrix, $W_{n}(\theta)=(n^{-1}\sum_{i=1}^{n}g(X_{i},\theta)g(X_{i},\theta)')^{-1}$, for the two-step GMM estimator. This does not affect the main result of the paper, though the resulting pseudo-true values are different. In practice, however, the uncentered weight matrix may not behave well under misspecification, because the elements of the uncentered weight matrix include bias terms of the moment function. See Hall (2000) for more discussion on the issue.}
\begin{equation}
W_{n}(\theta) = \left(n^{-1}\sum_{i=1}^{n}(g(X_{i},\theta)-g_{n}(\theta))(g(X_{i},\theta)-g_{n}(\theta))'\right)^{-1},
\end{equation}
and $g_{n}(\theta) = n^{-1}\sum_{i=1}^{n}g(X_{i},\theta)$. Suppress the dependence of $W_{n}$ on $\theta$ and write $W_{n}\equiv W_{n}(\hat{\theta}_{(1)})$. Under regularity conditions, the GMM estimators are consistent: $\hat{\theta}_{(1)}$ converges to a pseudo-true value $\theta_{0}(I)\equiv\theta_{0(1)}$, and $\hat{\theta}_{(2)}$ converges to a pseudo-true value $\theta_{0}(W)\equiv\theta_{0(2)}$. Under misspecification, $\theta_{0(1)}\neq\theta_{0(2)}$ in general. The probability limit of the weight matrix $W_{n}$ is $W = \left\{E[(g(X_{i},\theta_{0(1)})-g_{0(1)})(g(X_{i},\theta_{0(1)})-g_{0(1)})']\right\}^{-1}$, where $g_{0(j)}=Eg(X_{i},\theta_{0(j)})$ for $j=1,2$.

To further simplify notation, let $G(X_{i},\theta)=(\partial/\partial\theta')g(X_{i},\theta)$,
\begin{equation}
G_{0(j)}=EG(X_{i},\theta_{0(j)}),\hspace{1em}G_{0(j)}^{(2)}=E\left[\frac{\partial}{\partial\theta'}vec\left\{G(X_{i},\theta_{0(j)})\right\}\right],
\end{equation}
for $j=1,2$, and $L_{\theta}\times L_{\theta}$ matrices $H_{0(1)}=G_{0(1)}' G_{0(1)}+(g_{0(1)}'\otimes I_{L_{\theta}})G_{0(1)}^{(2)}$ and $H_{0(2)}=G_{0(2)}'W G_{0(2)}+(g_{0(2)}'W\otimes I_{L_{\theta}})G_{0(2)}^{(2)}$. Let
\begin{equation}
G_{n}(\theta)=n^{-1}\sum_{i=1}^{n}G(X_{i},\theta),\hspace{1em}G_{n}^{(2)}(\theta)=n^{-1}\sum_{i=1}^{n}\frac{\partial}{\partial\theta'}vec\left\{G(X_{i},\theta)\right\},
\end{equation}
$G_{n(j)}=G_{n}(\hat{\theta}_{(j)})$ for $j=1,2$, and $H_{n(1)}=G_{n(1)}' G_{n(1)}+(g_{n(1)}'\otimes I_{L_{\theta}})G_{n(1)}^{(2)}$ and $H_{n(2)}=G_{n(2)}'W_{n} G_{n(2)}+(g_{n(2)}'W_{n}\otimes I_{L_{\theta}})G_{n(2)}^{(2)}$. Let $\Omega_{1}$ and $\Omega_{2}$ denote positive-definite matrices such that
\begin{equation}
\sqrt{n}\left(
    \begin{array}{c}
      (g_{n}(\theta_{0(1)})-g_{0(1)}) \\
      (G_{n}(\theta_{0(1)})-G_{0(1)})'g_{0(1)} \\
    \end{array}
  \right)\rightarrow_{d}N\left(\mathbf{0}, \underset{(L_{g}+L_{\theta})\times(L_{g}+L_{\theta})}{\Omega_{1}}\right),
\end{equation}
and
\begin{equation}
\sqrt{n}\left(
    \begin{array}{c}
      (g_{n}(\theta_{0(2)})-g_{0(2)}) \\
      (G_{n}(\theta_{0(2)})-G_{0(2)})'Wg_{0(2)} \\
      (W_{n}-W)g_{0(2)} \\
    \end{array}
  \right)\rightarrow_{d}N\left(\mathbf{0}, \underset{(2L_{g}+L_{\theta})\times(2L_{g}+L_{\theta})}{\Omega_{2}}\right).
\end{equation}

To obtain the MR asymptotic covariance matrix for the GMM estimator, I use Theorems 1 and 2 of Hall and Inoue (2003):
\begin{equation}
\sqrt{n}(\hat{\theta}_{(j)}-\theta_{0(j)})\rightarrow_{d}N(0,\Sigma_{MR(j)}),
\end{equation}
where $\Sigma_{MR(j)}=H_{0(j)}^{-1}V_{j}H_{0(j)}^{-1'}$, for $j=1,2,$
\begin{eqnarray}
 V_{1}&=&\left[
       \begin{array}{cc}
         G_{0(1)}' & I_{L_{\theta}} \\
       \end{array}
     \right]\Omega_{1}\left[
       \begin{array}{cc}
         G_{0(1)}' & I_{L_{\theta}} \\
       \end{array}
     \right]',\\
\nonumber V_{2}&=&\left[
       \begin{array}{ccc}
         G_{0(2)}'W & I_{L_{\theta}} & G_{0(2)}'\\
       \end{array}
     \right]\Omega_{2}\left[
       \begin{array}{ccc}
         G_{0(2)}'W & I_{L_{\theta}} & G_{0(2)}'\\
       \end{array}
     \right]'.
\end{eqnarray}
Under correct specifications, $\Sigma_{MR(1)}$ and $\Sigma_{MR(2)}$ reduce to the standard asymptotic covariance matrices of the GMM estimators, $\Sigma_{C(1)}$ and $\Sigma_{C(2)}$ respectively, where
\begin{equation}
\Sigma_{C(1)} = (G_{0}'G_{0})^{-1}G_{0}'\Omega_{C}G_{0}(G_{0}'G_{0})^{-1},\hspace{5mm}\Sigma_{C(2)} = (G_{0}'\Omega_{C}^{-1}G_{0})^{-1},
\end{equation}
$G_{0}=EG(X_{i},\theta_{0})$, $\Omega_{C} = E[g(X_{i},\theta_{0})g(X_{i},\theta_{0})']$, and $\theta_{0}$ satisfies $Eg(X_{i},\theta_{0})=0$.

A consistent estimator of $\Sigma_{MR(j)}$ is $\hat{\Sigma}_{MR(j)} = H_{n(j)}^{-1}V_{n(j)}H_{n(j)}^{-1'}$ for $j=1,2,$ where
\begin{eqnarray}
 V_{n(1)}&=&\left[
       \begin{array}{cc}
         G_{n(1)}' & I_{L_{\theta}} \\
       \end{array}
     \right]\Omega_{n(1)}\left[
       \begin{array}{cc}
         G_{n(1)}' & I_{L_{\theta}} \\
       \end{array}
     \right]',\\
\nonumber V_{n(2)}&=&\left[
       \begin{array}{ccc}
         G_{n(2)}'W_{n} & I_{L_{\theta}} & G_{n(2)}'\\
       \end{array}
     \right]\Omega_{n(2)}\left[
       \begin{array}{ccc}
         G_{n(2)}'W_{n} & I_{L_{\theta}} & G_{n(2)}'\\
       \end{array}
     \right]',
\end{eqnarray}
and $\Omega_{n(j)}$ is a consistent estimator of $\Omega_{j}$, with the population moments replaced by the sample moments. In particular,
\begin{eqnarray}
 \Omega_{n(1)}&=&n^{-1}\sum_{i=1}^{n}\left(
                                  \begin{array}{c}
                                    g(X_{i},\hat{\theta}_{(1)})-g_{n(1)} \\
                                    (G(X_{i},\hat{\theta}_{(1)})-G_{n(1)})'g_{n(1)} \\
                                  \end{array}
                                \right)
                                \left(
                                  \begin{array}{c}
                                    g(X_{i},\hat{\theta}_{(1)})-g_{n(1)} \\
                                    (G(X_{i},\hat{\theta}_{(1)})-G_{n(1)})'g_{n(1)} \\
                                  \end{array}
                                \right)',\\
\nonumber \Omega_{n(2)}&=&n^{-1}\sum_{i=1}^{n}\left(
                                  \begin{array}{c}
                                    g(X_{i},\hat{\theta}_{(2)})-g_{n(2)} \\
                                    (G(X_{i},\hat{\theta}_{(2)})-G_{n(2)})'W_{n}g_{n(2)} \\
                                    W_{i}g_{n(2)} \\
                                  \end{array}
                                \right)
                                \left(
                                  \begin{array}{c}
                                    g(X_{i},\hat{\theta}_{(2)})-g_{n(2)} \\
                                    (G(X_{i},\hat{\theta}_{(2)})-G_{n(2)})'W_{n}g_{n(2)} \\
                                    W_{i}g_{n(2)} \\
                                  \end{array}
                                \right)',
\end{eqnarray}
where\footnote{Note that $W_{n}-W=-W(W_{n}^{-1}-W^{-1})W_{n}$.}
\begin{equation}
W_{i} = -W_{n}\cdot\left((g(X_{i},\hat{\theta}_{(1)})-g_{n}(\hat{\theta}_{(1)}))(g(X_{i},\hat{\theta}_{(1)})-g_{n}(\hat{\theta}_{(1)}))'-W_{n}^{-1}\right)\cdot W_{n}.
\end{equation}
The diagonal elements of the covariance estimator $\hat{\Sigma}_{MR(j)}$ for $j=1,2$ are the Hall-Inoue variance estimators. In practice, the estimation of the MR covariance matrices does not involve much complication. What we need to calculate additionally is the second derivative of the moment function.

Let $\theta_{k}$, $\theta_{0(j),k}$, and $\hat{\theta}_{(j),k}$ denote the $k$th elements of $\theta$, $\theta_{0(j)}$, and $\hat{\theta}_{(j)}$ respectively. Let $(\hat{\Sigma}_{MR(j)})_{kk}$ denote the $(k,k)$th element of $\hat{\Sigma}_{MR(j)}$. The $t$ statistic for testing the null hypothesis $H_{0}:\theta_{k}=\theta_{0(j),k}$ is
\begin{equation}
T_{MR(j)} = \frac{\hat{\theta}_{(j),k}-\theta_{0(j),k}}{\sqrt{(\hat{\Sigma}_{MR(j)})_{kk}/n}},
\end{equation}
where $j=1$ for the one-step GMM estimator and $j=2$ for the two-step GMM estimator. $T_{MR(j)}$ is robust to misspecification because it is asymptotically standard normal under $H_{0}$, without assuming the correct model. $T_{MR(j)}$ is different from the conventional $t$ statistic, because $\hat{\Sigma}_{C(j)}\neq\hat{\Sigma}_{MR(j)}$ in general even under correct specification, for $j=1,2$.\footnote{Applied researchers may be interested in the choice between $T_{MR(1)}$ and $T_{MR(2)}$. However, it is hard to compare them because (i) $\hat{\theta}_{(1)}$ and $\hat{\theta}_{(2)}$ have different probability limits, and (ii) efficiency gain of the two-step GMM does not hold anymore under misspecification. Nevertheless, comparing $T_{C(j)}$ and $T_{MR(j)}$ would be helpful in practice, where $T_{C(j)}$ is the conventional $t$ statistic studentized with $\hat{\Sigma}_{C(j)}$ for $j=1,2$. For example, one might want to use $T_{C(1)}$ instead of $T_{C(2)}$ to avoid a potential finite sample bias in the two-step GMM. In this case, it is recommended to calculate $T_{MR(1)}$ and compare it with $T_{C(1)}$. In general, $T_{MR(1)}$ is a better choice than $T_{C(1)}$ because it is robust to misspecification while it is not necessarily less powerful than $T_{C(1)}$ (see Section 7). A similar argument applies to $T_{C(2)}$ and $T_{MR(2)}$.} Note that $\hat{\Sigma}_{C(j)}$ is a consistent estimator for $\Sigma_{C(j)}$, the asymptotic covariance matrix under correct specification for $j=1,2$.

The MR bootstrap described in the next section achieves asymptotic refinements over the MR asymptotic $t$ test and CI, rather than the conventional non-robust ones. Define the MR asymptotic $t$ test and CI as follows. The symmetric two-sided $t$ test with asymptotic significance level $\alpha$ rejects $H_{0}$ if $|T_{MR(j)}|>z_{\alpha/2}$, where $z_{\alpha/2}$ is the $1-\alpha/2$ quantile of the standard normal distribution. The corresponding CI for $\theta_{0(j),k}$ with asymptotic confidence level $100(1-\alpha)\%$ is $CI_{MR(j)}=[\hat{\theta}_{(j),k}\pm z_{\alpha/2}\sqrt{(\hat{\Sigma}_{MR(j)})_{kk}/n}]$, $j=1,2$. The error in the rejection probability of the $t$ test with $z_{\alpha/2}$ and coverage probability of $CI_{MR(j)}$ is $O(n^{-1})$: Under $H_{0}$,
$P\left(|T_{MR(j)}|>z_{\alpha/2}\right)=\alpha+O(n^{-1}) \mbox{ and } P\left(\theta_{0(j),k}\in CI_{MR(j)}\right)=1-\alpha+O(n^{-1}),$
for $j=1,2$.

%%%%%%%%%%%%%%%%%%%%%%%%%%%%%%%%%%%%%%%%%%%%%%%%%%%%%%%%%%%%%%%%%%%%%%%%
\section{The Misspecification-Robust Bootstrap}
\label{MRB}
%%%%%%%%%%%%%%%%%%%%%%%%%%%%%%%%%%%%%%%%%%%%%%%%%%%%%%%%%%%%%%%%%%%%%%%%

The nonparametric iid bootstrap is implemented by sampling $X_{1}^{*},\cdots,X_{n}^{*}$ randomly with replacement from the sample $X_{1},\cdots,X_{n}$.

The bootstrap one-step GMM estimator, $\hat{\theta}_{(1)}^{*}$ solves:
\begin{equation}
\min_{\theta\in\Theta}J_{n}^{*}(\theta,I_{L_{g}})= \left(n^{-1}\sum_{i=1}^{n}g(X_{i}^{*},\theta)\right)'\left(n^{-1}\sum_{i=1}^{n}g(X_{i}^{*},\theta)\right),
\end{equation}
and the bootstrap two-step GMM estimator $\hat{\theta}_{(2)}^{*}$ solves
\begin{equation}
 \min_{\theta\in\Theta}J_{n}^{*}(\theta,W_{n}^{*}(\hat{\theta}^{*}_{(1)}))= \left(n^{-1}\sum_{i=1}^{n}g(X_{i}^{*},\theta)\right)'W_{n}^{*}(\hat{\theta}^{*}_{(1)})\left(n^{-1}\sum_{i=1}^{n}g(X_{i}^{*},\theta)\right),
\end{equation}
where
\begin{equation}
W_{n}^{*}(\theta) = \left(n^{-1}\sum_{i=1}^{n}(g(X_{i}^{*},\theta)-g_{n}^{*}(\theta))(g(X_{i}^{*},\theta)-g_{n}^{*}(\theta))'\right)^{-1},
\end{equation}
and $g_{n}^{*}(\theta) = n^{-1}\sum_{i=1}^{n}g(X_{i}^{*},\theta)$. Suppress the dependence of $W_{n}^{*}$ on $\theta$ and write $W_{n}^{*}\equiv W_{n}^{*}(\hat{\theta}_{(1)}^{*})$. To further simplify notation, let
\begin{equation}
G_{n}^{*}(\theta)=n^{-1}\sum_{i=1}^{n}\frac{\partial}{\partial\theta'}g(X_{i}^{*},\theta),\hspace{1em}G_{n}^{(2)*}(\theta)=n^{-1}\sum_{i=1}^{n}\frac{\partial}{\partial\theta'}vec\left\{\frac{\partial}{\partial\theta'}g(X_{i}^{*},\theta)\right\},
\end{equation}
$G_{n(j)}^{*}=G_{n}^{*}(\hat{\theta}^{*}_{(j)})$ for $j=1,2$, and $H_{n(1)}^{*}=G_{n(1)}^{*'} G_{n(1)}^{*}+(g_{n(1)}^{*'}\otimes I_{L_{\theta}})G_{n(1)}^{(2)*}$ and $H_{n(2)}^{*}=G_{n(2)}^{*'}W_{n}^{*}G_{n(2)}^{*}+(g_{n(2)}^{*'}W_{n}^{*}\otimes I_{L_{\theta}})G_{n(2)}^{(2)*}$.

The bootstrap version of the robust covariance matrix estimator $\hat{\Sigma}_{MR(j)}$ is $\hat{\Sigma}_{MR(j)}^{*} = H_{n(j)}^{*-1}V_{n(j)}^{*}H_{n(j)}^{*-1'}$ for $j=1,2,$ where
\begin{eqnarray}
 V_{n(1)}^{*}&=&\left[
       \begin{array}{cc}
         G_{n(1)}^{*'} & I_{L_{g}} \\
       \end{array}
     \right]\Omega_{n(1)}^{*}\left[
       \begin{array}{cc}
         G_{n(1)}^{*'} & I_{L_{g}} \\
       \end{array}
     \right]',\\
\nonumber V_{n(2)}^{*}&=&\left[
       \begin{array}{ccc}
         G_{n(2)}^{*'}W_{n}^{*} & I_{L_{g}} & G_{n(2)}^{*'}\\
       \end{array}
     \right]\Omega_{n(2)}^{*}\left[
       \begin{array}{ccc}
         G_{n(2)}^{*'}W_{n}^{*} & I_{L_{g}} & G_{n(2)}^{*'}\\
       \end{array}
     \right]',
\end{eqnarray}
and $\Omega_{n(j)}^{*}$ is constructed by replacing the sample moments in $\Omega_{n(j)}$ with the bootstrap sample moments. In particular,
\begin{eqnarray}
 \Omega_{n(1)}^{*}&=&n^{-1}\sum_{i=1}^{n}\left(
                                  \begin{array}{c}
                                    g(X_{i}^{*},\hat{\theta}_{(1)}^{*})-g_{n(1)}^{*} \\
                                    (G(X_{i}^{*},\hat{\theta}_{(1)}^{*})-G_{n(1)}^{*})'g_{n(1)}^{*} \\
                                  \end{array}
                                \right)
                                \left(
                                  \begin{array}{c}
                                    g(X_{i}^{*},\hat{\theta}_{(1)}^{*})-g_{n(1)}^{*} \\
                                    (G(X_{i}^{*},\hat{\theta}_{(1)}^{*})-G_{n(1)}^{*})'g_{n(1)}^{*} \\
                                  \end{array}
                                \right)',\\
\nonumber \Omega_{n(2)}^{*}&=&n^{-1}\sum_{i=1}^{n}\left(
                                  \begin{array}{c}
                                    g(X_{i}^{*},\hat{\theta}_{(2)}^{*})-g_{n(2)}^{*} \\
                                    (G(X_{i}^{*},\hat{\theta}_{(2)}^{*})-G_{n(2)}^{*})'W_{n}^{*}g_{n(2)}^{*} \\
                                    W_{i}^{*}g_{n(2)}^{*} \\
                                  \end{array}
                                \right)
                                \left(
                                  \begin{array}{c}
                                    g(X_{i}^{*},\hat{\theta}_{(2)}^{*})-g_{n(2)}^{*} \\
                                    (G(X_{i}^{*},\hat{\theta}_{(2)}^{*})-G_{n(2)}^{*})'W_{n}^{*}g_{n(2)}^{*} \\
                                    W_{i}^{*}g_{n(2)}^{*} \\
                                  \end{array}
                                \right)',
\end{eqnarray}
where
\begin{equation}
W_{i}^{*} = -W_{n}^{*}\cdot\left((g(X_{i}^{*},\hat{\theta}_{(1)}^{*})-g_{n}^{*}(\hat{\theta}_{(1)}^{*}))(g(X_{i}^{*},\hat{\theta}_{(1)}^{*})-g_{n}^{*}(\hat{\theta}_{(1)}^{*}))'-W_{n}^{*-1}\right)\cdot W_{n}^{*}.
\end{equation}

The MR bootstrap $t$ statistic is
\begin{equation}
T_{MR(j)}^{*} = \frac{\hat{\theta}_{(j),k}^{*}-\hat{\theta}_{(j),k}}{\sqrt{(\hat{\Sigma}^{*}_{MR(j)})_{kk}/n}},
\end{equation}
for $j=1,2$. Let $z^{*}_{|T_{MR(j)}|,\alpha}$ denote the $1-\alpha$ quantile of $|T_{MR(j)}^{*}|$, $j=1,2$. Following Andrews (2002), we define $z^{*}_{|T_{MR(j)}|,\alpha}$ to be a value that minimizes $|P^{*}(|T_{MR(j)}^{*}|\leq z)-(1-\alpha)|$ over $z\in \mathbf{R}$, since the distribution of $|T_{MR(j)}^{*}|$ is discrete. The symmetric two-sided bootstrap $t$ test of $H_{0}:\theta_{k}=\theta_{0(j),k}$ versus $H_{1}:\theta_{k}\neq \theta_{0(j),k}$ rejects if $|T_{MR(j)}|>z^{*}_{|T_{MR(j)}|,\alpha}$, $j=1,2$, and this test is of asymptotic significance level $\alpha$. The $100(1-\alpha)\%$ symmetric percentile-$t$ interval for $\theta_{0(j),k}$ is, for $j=1,2$,
\begin{equation}
CI_{MR(j)}^{*}=\left[\hat{\theta}_{(j),k}\pm z^{*}_{|T_{MR(j)}|,\alpha}\sqrt{(\hat{\Sigma}_{MR(j)})_{kk}/n}\right].
\end{equation}

The MR bootstrap $t$ statistic differs from the recentered bootstrap $t$ statistic. First, unlike the Hall-Horowitz bootstrap, the MR bootstrap GMM estimator is calculated from the original moment function with the bootstrap sample. Second, the Hall-Inoue variance estimator is used to construct the bootstrap $t$ statistic. In the recentered bootstrap, the conventional variance estimator of Hansen (1982) is used.

%%%%%%%%%%%%%%%%%%%%%%%%%%%%%%%%%%%%%%%%%%%%%%%%%%%%%%%%%
\section{Main Result}
\label{Main}
%%%%%%%%%%%%%%%%%%%%%%%%%%%%%%%%%%%%%%%%%%%%%%%%%%%%%%%%%

%%%%%%%%%%%%%%%%%%%%%%%%%%%%%%%%%%%%%%%%%%%%%%%%%%%%%%%%%
\subsection{Assumptions}
%%%%%%%%%%%%%%%%%%%%%%%%%%%%%%%%%%%%%%%%%%%%%%%%%%%%%%%%%

The assumptions are analogous to those of Hall and Horowitz (1996) and Andrews (2002). The main difference is that I do not assume correct model specification. If the model is misspecified, then the probability limits of the one-step and the two-step GMM estimators are different. Thus, we need to distinguish $\theta_{0(1)}$ from $\theta_{0(2)}$, the probability limit of $\hat{\theta}_{(1)}$ and $\hat{\theta}_{(2)}$, respectively. The assumptions are modified to hold for both pseudo-true values. If the model happens to be correctly specified, then the pseudo-true values become identical.

Let $f(X_{i},\theta)$ denote the vector containing the unique components of $g(X_{i},\theta)$ and $g(X_{i},\theta)g(X_{i},\theta)'$, and their derivatives through order $d_{1}\geq 6$ with respect to $\theta$. Let $(\partial^{m}/\partial\theta^{m})g(X_{i},\theta)$ and $(\partial^{m}/\partial\theta^{m})f(X_{i},\theta)$ denote the vectors of partial derivatives with respect to $\theta$ of order $m$ of $g(X_{i},\theta)$ and $f(X_{i},\theta)$, respectively.

\begin{assumption}
$X_{i},i=1,2,...$ are iid.
\label{A1}
\end{assumption}

\begin{assumption}\
(a) $\Theta$ is compact and $\theta_{0(1)}$ and $\theta_{0(2)}$ are interior points of $\Theta$.\\
(b) $\hat{\theta}_{(1)}$ and $\hat{\theta}_{(2)}$ minimize $J_{n}(\theta,I_{L_{g}})$ and $J_{n}(\theta,W_{n})$ over $\theta\in\Theta$, respectively; $\theta_{0(1)}$ and $\theta_{0(2)}$ are the pseudo-true values that uniquely minimize $J(\theta,I_{L_{g}})$ and $J(\theta,W)$ over $\theta\in\Theta$, respectively; for some function $C_{g}(x)$, $\|g(x,\theta_{1})-g(x,\theta_{2})\|<C_{g}(x)\|\theta_{1}-\theta_{2}\|$ for all $x$ in the support of $X_{1}$ and all $\theta_{1},\theta_{2}\in\Theta$; and $EC_{g}^{q_{1}}(X_{1})<\infty$ and $E\|g(X_{1},\theta)\|^{q_{1}}<\infty$ for all $\theta\in\Theta$ for all $0<q_{1}<\infty$.
\label{A2}
\end{assumption}

\begin{assumption}
The followings hold for $j=1,2$.\\
(a) $\Omega_{j}$ is positive definite.\\
(b) $H_{0(j)}$ is nonsingular and $G_{0(j)}$ is full rank $L_{\theta}$.\\
(c) $g(x,\theta)$ is $d=d_{1}+d_{2}$ times differentiable with respect to $\theta$ on $N_{0(j)}$, where $N_{0(j)}$ is some neighborhood of $\theta_{0(j)}$, for all $x$ in the support of $X_{1}$, where $d_{1}\geq6$ and $d_{2}\geq5$.\\
(d) There is a function $C_{\partial f}(X_{1})$ such that $\|(\partial^{m}/\partial\theta^{m})f(X_{1},\theta)-(\partial^{m}/\partial\theta^{m})f(X_{1},\theta_{0(j)})\|\leq C_{\partial f}(X_{1})\|\theta-\theta_{0(j)}\|$ for all $\theta\in N_{0(j)}$ for all $m=0,...,d_{2}$.\\
(e) $EC^{q_{2}}_{\partial f}(X_{1})<\infty$ and $E\|(\partial^{m}/\partial\theta^{m})f(X_{1},\theta_{0(j)})\|^{q_{2}}\leq C_{f}<\infty$ for all $m=0,...,d_{2}$ for some constant $C_{f}$ (that may depend on $q_{2}$) and all $0<q_{2}<\infty$.\\
(f) $f(X_{1},\theta_{0(j)})$ is once differentiable with respect to $X_{1}$ with uniformly continuous first derivative.
\label{A3}
\end{assumption}

\begin{assumption}
For $t\in\mathbf{R}^{dim(f)}$ and $j=1,2$, $\limsup_{\|t\|\rightarrow\infty}\left|E\left(\exp(it'f(X_{1},\theta_{0(j)}))\right)\right|<1,$ where $i=\sqrt{-1}$.
\label{A4}
\end{assumption}

Assumption \ref{A1} says that we restrict our attention to iid sample. Hall and Horowitz (1996) and Andrews (2002) deal with dependent data. I focus on iid sample and nonparametric iid bootstrap to emphasize the role of the Hall-Inoue variance estimator in implementing the MR bootstrap without recentering and to avoid the complications arising when constructing blocks to deal with dependent data. For example, the Hall-Horowitz bootstrap needs an additional correction factor as well as recentering for dependent data. The correction factor would also be needed in implementing the MR bootstrap for dependent data. I do not investigate this issue further in this paper.

Assumptions \ref{A2}-\ref{A3} are similar to Assumptions 2-3 of Andrews (2002), except that I eliminate the correct model assumption. In particular, I relax Assumption 2 of Hall and Horowitz (1996) and Assumption 2(b)(i) of Andrews (2002). The moment conditions in Assumptions \ref{A2}-\ref{A3} are not primitive, but they lead to simpler results as in Andrews (2002). Assumption \ref{A4} is the standard Cram\'{e}r condition for iid sample, that is needed to get Edgeworth expansions.

%%%%%%%%%%%%%%%%%%%%%%%%%%%%%%%%%%%%%%%%%%%%%%%%%%%%%%%%%%%%%%%%%%%%%%%%%%%%%%%
\subsection{Asymptotic Refinements of the Misspecification-Robust Bootstrap}
%%%%%%%%%%%%%%%%%%%%%%%%%%%%%%%%%%%%%%%%%%%%%%%%%%%%%%%%%%%%%%%%%%%%%%%%%%%%%%%

Theorem \ref{T1} shows that the MR bootstrap symmetric two-sided $t$ test has rejection probability that is correct up to $O(n^{-2})$, and the same magnitude of convergence holds for the MR bootstrap symmetric percentile-$t$ interval. This result extends the results of Theorem 3 of Hall and Horowitz (1996) and Theorem 2(c) of Andrews (2002), because their results hold only under correctly specified models. In other words, the following Theorem establishes that the MR bootstrap achieves the same magnitude of asymptotic refinements with the existing bootstrap procedures, without assuming the correct model and without recentering.

\begin{theorem}
Suppose Assumptions \ref{A1}-\ref{A4} hold. Under $H_{0}:\theta_{k}=\theta_{0(j),k}$, for $j=1,2,$
$$P(|T_{MR(j)}|>z^{*}_{|T_{MR(j)}|,\alpha})=\alpha+O(n^{-2})\hspace{5mm}\mbox{ or }\hspace{5mm}P(\theta_{0(j),k}\in CI_{MR(j)}^{*})=1-\alpha+O(n^{-2}),$$
where $z^{*}_{|T_{MR(j)}|,\alpha}$ is the $1-\alpha$ quantile of the distribution of $|T_{MR(j)}^{*}|$.
\label{T1}
\end{theorem}

Since $P\left(|T_{MR(j)}|>z_{\alpha/2}\right)=\alpha+O(n^{-1})$, the bootstrap critical value has a reduction in the error of rejection probability by a factor of $n^{-1}$ for symmetric two-sided $t$ tests. The symmetric percentile-$t$ interval is formulated by the symmetric two-sided $t$ test, and the CI also has a reduction in the error of coverage probability by a factor of $n^{-1}$.

We note that neither asymptotic refinements nor first-order validity for the $J$ test are established in Theorem \ref{T1}. The MR bootstrap is implemented with a misspecified moment condition in the sample, $E^{*}g(X_{i}^{*},\hat{\theta})\neq 0$, where $E^{*}$ is the expectation over the bootstrap sample. Thus, the distribution of the MR bootstrap $J$ statistic does not consistently approximate that of the sample $J$ statistic under the null hypothesis, which is $Eg(X_{i},\theta_{0})=0$.

The proof of the Theorem proceeds by showing that the misspecification-robust $t$ statistic studentized with the Hall-Inoue variance estimator can be approximated by a smooth function of sample moments. Once we establish that the approximation is close enough, we can use the result of Edgeworth expansions for a smooth function in Hall (1992). The proof extensively follows those of Hall and Horowitz (1996) and Andrews (2002). The differences are that I allow for distinct probability limits of the one-step and the two-step GMM estimators, and that no special bootstrap version of the test statistic is needed for the MR bootstrap. Indeed, the recentering creates more complication than it seems even under correct specification, because $\hat{\theta}_{(1)}\neq\hat{\theta}_{(2)}$ in general, which in turn implies that there are two (pseudo-)true values in the bootstrap world. This issue is not explicitly explained in Hall and Horowitz (1996) and Andrews (2002). In contrast, I explicitly distinguish the pseudo-true values in the bootstrap world as well as in the population, which makes the proof given in this paper more straightforward than theirs.

%%%%%%%%%%%%%%%%%%%%%%%%%%%%%%%%%%%%%%%%%%%%%%%%%%%%%%%%%%%%%%%
\section{Monte Carlo Experiments}
\label{MC}
%%%%%%%%%%%%%%%%%%%%%%%%%%%%%%%%%%%%%%%%%%%%%%%%%%%%%%%%%%%%%%%

In this section, I compare the actual finite sample coverage probabilities of the asymptotic and bootstrap CI's under correct specification and misspecification.

The conventional asymptotic CI with coverage probability $100(1-\alpha)\%$ is
\begin{equation}
CI_{C}=\left[\hat{\theta}\pm z_{\alpha/2}\sqrt{\hat{\Sigma}_{C}/n}\right],
\end{equation}
where $z_{\alpha/2}$ is the $1-\alpha/2$th quantile of the standard normal distribution. The MR asymptotic CI using the Hall-Inoue variance estimator with coverage probability $100(1-\alpha)\%$ is
\begin{equation}
CI_{MR}=\left[\hat{\theta}\pm z_{\alpha/2}\sqrt{\hat{\Sigma}_{MR}/n}\right].
\end{equation}
The only difference between $CI_{MR}$ and $CI_{C}$ is the choice of the variance estimator. Under correct model specification, both the asymptotic CI's have coverage probability $100(1-\alpha)\%$ asymptotically and the error in the coverage probability is $O(n^{-1})$. Under misspecification, $CI_{MR}$ still provides asymptotically correct coverage, but $CI_{C}$ does not because $\hat{\Sigma}_{C}$ is inconsistent.

The Hall-Horowitz and the Brown-Newey bootstrap CI's with coverage probability $100(1-\alpha)\%$ are given by
\begin{eqnarray}
 CI_{HH}^{*}&=&\left[\hat{\theta}\pm z_{|T_{HH}|,\alpha}^{*}\sqrt{\hat{\Sigma}_{C}/n}\right],\\
 CI_{BN}^{*}&=&\left[\hat{\theta}\pm z_{|T_{BN}|,\alpha}^{*}\sqrt{\hat{\Sigma}_{C}/n}\right],
\end{eqnarray}
where $z_{|T_{HH}|,\alpha}^{*}$ and $z_{|T_{BN}|,\alpha}^{*}$ are the $1-\alpha$th quantiles of the bootstrap distribution of the absolute value of the $t$ statistic based on the Hall-Horowitz bootstrap and the Brown-Newey bootstrap, respectively. Both the recentered bootstrap CI's are expected to perform better than $CI_{C}$ under correct specification. However, similar to $CI_{C}$, they do not provide asymptotically correct coverage under misspecification.

The MR bootstrap CI with coverage probability $100(1-\alpha)\%$ is:
\begin{equation}
CI_{MR}^{*}=\left[\hat{\theta}\pm z_{|T_{MR}|,\alpha}^{*}\sqrt{\hat{\Sigma}_{MR}/n}\right],
\end{equation}
where $z_{|T_{MR}|,\alpha}^{*}$ is the $1-\alpha$th quantile of the MR bootstrap distribution of the absolute value of the $t$ statistic. $CI_{MR}^{*}$ is expected to perform better than $CI_{MR}$ regardless of misspecification by Theorem \ref{T1}.

%%%%%%%%%%%%%%%%%%%%%%%%%%%%%%%%%%%%%%%%%%%%%%%%%%%%%
\subsection{Example 1: Combining Data Sets}
%%%%%%%%%%%%%%%%%%%%%%%%%%%%%%%%%%%%%%%%%%%%%%%%%%%%%
Suppose that we observe $X_{i}=(Y_{i},Z_{i})'\in\mathbb{R}^{2}$, $i=1,...n$, and we have an econometric model based on $Z_{i}$ with a moment function $g_{1}(Z_{i},\theta)$, where $\theta$ is a parameter of interest. Also, suppose that we know the mean (or other population information) of $Y_{i}$. If $Y_{i}$ and $Z_{i}$ are correlated, we can exploit the known information on $EY_{i}$ to get more accurate estimates of $\theta$. This situation is common in survey sampling: A sample survey consists of a random sample from some population and aggregate statistics from the same population. Imbens and Lancaster (1994) and Hellerstein and Imbens (1999) show how to efficiently combine data sets and make an inference. For more examples, see Imbens (2002) and Section 3.10 of Owen (2001).

Let $g_{1}(Z_{i},\theta)=Z_{i}-\theta$, so that the parameter of interest is the mean of $Z_{i}$. Without the knowledge on $EY_{i}$, the natural estimator is the sample mean of $Z_{i}$. If an additional information, $EY_{i}=0$, is available, then we form the moment function as
\begin{equation}
g(X_{i},\theta)=\left(
                    \begin{array}{c}
                      Y_{i} \\
                      Z_{i}-\theta \\
                    \end{array}
                  \right).
\end{equation}
Since the number of moment restrictions ($L_{g}=2$) is greater than that of the parameter ($L_{\theta}=1$), the model is overidentified and we can use GMM estimators to estimate $\theta$. If the assumed mean of $Y$ is not true, i.e., $EY_{i}\neq0$, then the model is misspecified because there is no $\theta$ that satisfies $Eg(X_{i},\theta)=0$.

The one-step GMM estimator solving \eqref{1gmm} is given by $\hat{\theta}_{(1)}=\bar{Z}\equiv n^{-1}\sum_{i=1}^{n}Z_{i}$. The two-step GMM estimator solving \eqref{2gmm} and the pseudo-true value are given by \begin{equation}
\hat{\theta}_{(2)}=\bar{Z}-\frac{\widehat{Cov}(Y_{i},Z_{i})}{\widehat{Var}(Y_{i})}\bar{Y}\rightarrow_{p}\theta_{0(2)}=EZ_{i}-\frac{Cov(Y_{i},Z_{i})}{Var(Y_{i})}EY_{i},
\end{equation}
where $\widehat{Var}(Y_{i})=n^{-1}\sum_{i=1}^{n}(Y_{i}-\bar{Y})^{2}$ and $\widehat{Cov}(Y_{i},Z_{i})=n^{-1}\sum_{i=1}^{n}(Y_{i}-\bar{Y})(Z_{i}-\bar{Z})$. Note that the pseudo-true value reduces to $\theta_{0(2)}=EZ_{i}$ when $EY_{i}=0$, i.e., the model is correctly specified.

The conventional asymptotic variance of $\hat{\theta}_{(2)}$ is $\Sigma_{C(2)}=(G_{0}'\Omega_{C}^{-1}G_{0})^{-1}$. The MR asymptotic variance of $\hat{\theta}_{(2)}$ is $\Sigma_{MR(2)}$, where the formula for $\Sigma_{MR(2)}$ is given in the previous section. Note that $\Sigma_{C(2)}$ is a special case of $\Sigma_{MR(2)}$ imposing no misspecification. The following example makes this case clear. Consider a simple data generating process (DGP)
\begin{equation}
\left(
    \begin{array}{c}
      Y_{i} \\
      Z_{i} \\
    \end{array}
  \right)\sim N\left(\left(
                       \begin{array}{c}
                         \delta \\
                         0 \\
                       \end{array}
                     \right), \left(
                                \begin{array}{cc}
                                  1 & \rho \\
                                  \rho & 1 \\
                                \end{array}
                              \right)\right),
\label{OrgDGP}
\end{equation}
where $0<\rho<1$ is a correlation between $Y_{i}$ and $Z_{i}$, and $(Y_{i},Z_{i})'$ is iid. The assumed mean of $Y_{i}$, zero, may not equal to the true value, $\delta$. Therefore, $\delta$ measures a degree of misspecification. As $\delta$ deviates farther from zero, the degree of misspecification becomes larger. The pseudo-true value is $\theta_{0(2)}=-\rho\delta$, and the asymptotic variances $\Sigma_{C(2)}$ and $\Sigma_{MR(2)}$ are\footnote{See Supplementary Appendix for details about the calculation.}
\begin{equation}
\Sigma_{C(2)}=1-\rho^{2},\hspace{2em}\Sigma_{MR(2)}=(1-\rho^{2})(1+\delta^{2}).
\end{equation}
If the model is correctly specified, then using the additional information reduces the variance of the estimator by $\rho^{2}$, because the asymptotic variance of the sample mean $\bar{Z}$ is $Var(Z_{i})=1$. However, this reduction may not occur when the additional information is misspecified, and furthermore, the conventional variance estimator is inconsistent for the true asymptotic variance, $\Sigma_{MR(2)}$. In contrast, the Hall-Inoue variance estimator is consistent for the true asymptotic variance regardless of misspecification.

To better compare the coverage probabilities of the CI's, I modify the DGP \eqref{OrgDGP}:
\begin{equation}
\left(
    \begin{array}{c}
      Y_{i} \\
      Z_{i}^{0} \\
    \end{array}
  \right)\sim N\left(\left(
                       \begin{array}{c}
                         \delta \\
                         0 \\
                       \end{array}
                     \right), \left(
                                \begin{array}{cc}
                                  1 & \rho \\
                                  \rho & 1 \\
                                \end{array}
                              \right)\right),\hspace{1em}Z_{i} = e^{\sigma Z_{i}^{0}}-e^{\sigma^{2}/2},
\label{DGP1}
\end{equation}
where $\sigma$ is a shape parameter.\footnote{Unreported simulation results based on the DGP \eqref{OrgDGP} are similar to the reported one, although the size distortion of the asymptotic CI's are less severe.} In this case, $Z_{i}$ has a shifted log-normal distribution, and the mean and the variance are 0 and $(e^{\sigma^{2}}-1)e^{\sigma^{2}}$, respectively. Estimating the mean of $Z_{i}$ is a common problem in economics, as many economic data are well approximated by log-normal distributions. The information on the mean of $Y_{i}$ is assumed to be relatively accurate, but may not be exact, which is the source of misspecification.

Table \ref{table_ex1} shows the coverage probabilities of 90\% and 95\% CI's based on the two-step GMM estimator, $\hat{\theta}_{(2)}$, when $\rho=0.5$ and $\sigma=1.5$ in \eqref{DGP1}. The number of Monte Carlo repetition (r) is 5,000, and the number of bootstrap replication (B) is 1,000. \textit{$J$ ($J^{*}$) at 5\%} denotes the actual rejection probabilities of the asymptotic and the Hall-Horowitz bootstrap $J$ test at 5\% level.

For a correctly specified model ($\delta=0$), the bootstrap CI's show better performance than the asymptotic CI's for $n=50$, $200$, and $1,000$. One might suspect that $CI_{MR}^{*}$ and $CI_{MR}$ may not work well compared to the conventional CI's under correct specification ($\delta=0$). Interestingly, $CI_{MR}^{*}$ works as good as $CI_{HH}^{*}$ and $CI_{BN}^{*}$, and $CI_{MR}$ works as good as $CI_{C}$ under correct specification. This implies that the two variance estimators $\hat{\Sigma}_{MR}$ and $\hat{\Sigma}_{C}$ do not differ much, but the difference is enough to achieve asymptotic refinements of the bootstrap without recentering. Since $\hat{\Sigma}_{MR}$ involves estimation of the fourth moment of the moment function $g(X_{i},\theta)$, rather than the second moment, $\hat{\Sigma}_{MR}$  may not work well if we consider more complicated nonlinear models and DGP's. Their relative performance under correct specification deserves more research.

\begin{figure}[p!]
    \centering
    \includegraphics[width=120mm]{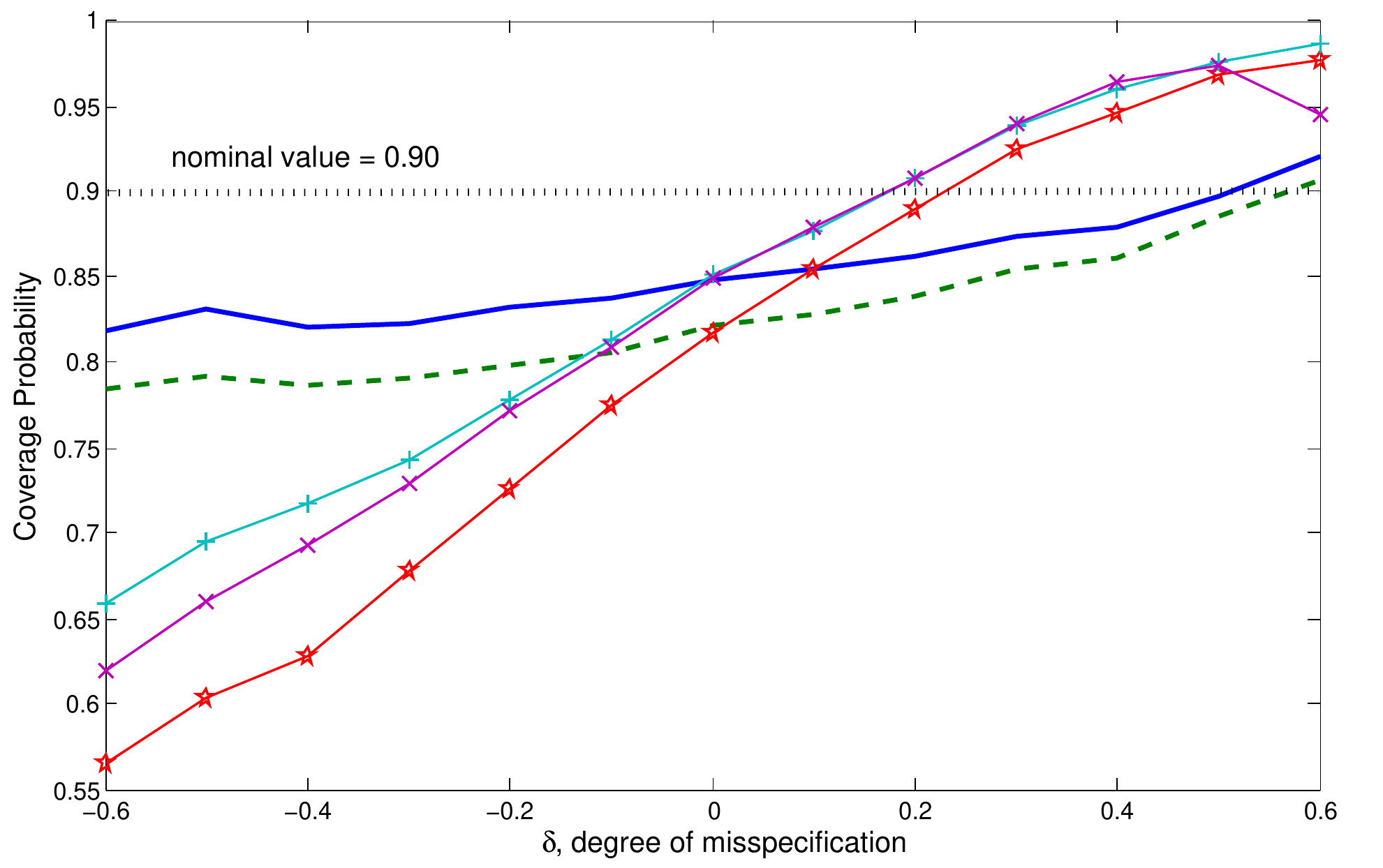}
    \caption{Coverage Probabilities of 90\% Confidence Intervals for $\theta_{0(2)}$ based on the Two-step GMM Estimator, $\hat{\theta}_{(2)}$, when $n=200$ in Example 1 DGP \eqref{DGP1}: $CI_{MR}^{*}$ (solid), $CI_{MR}$ (dashed), $CI_{C}$ (dashed with stars), $CI_{HH}^{*}$ (solid with +'s), $CI_{BN}^{*}$ (solid with x's)}
    \label{fig_ex1}
\end{figure}

For misspecified models ($\delta=-0.3,-0.6,0.6$), only $CI_{MR}^{*}$ and $CI_{MR}$ have asymptotically correct coverage. $CI_{MR}^{*}$ performs better than $CI_{MR}$ regardless of misspecification, which supports asymptotic refinements robust to misspecification. In contrast, the conventional asymptotic and bootstrap CI's are first-order invalid. Their coverage is either significantly lower (when $\delta=-0.6$) or significantly higher (when $\delta=0.6$) than the nominal coverage.\footnote{Under misspecification, the estimation of the empirical likelihood probabilities for the Brown-Newey bootstrap did not work well. For example, convergence failure occurred about 30\% of the Monte Carlo repetition when $\delta=0.6$ and $n=1,000$. If this happens, $CI_{BN}^{*}$ has a length zero, which trivially does not cover the pseudo-true value.} In particular, the result when $\delta=0.6$ implies that the conventional CI's may be neither asymptotically correct nor shorter in finite sample under misspecification. Figure \ref{fig_ex1} shows the coverage probabilities of the CI's when $n=200$ for different values of $\delta$, and also supports the findings above.

%%%%%%%%%%%%%%%%%%%%%%%%%%%%%%%%%%%%%%%%%%%%%%%%%%%%%%%%%%%%%%%%%%
\subsection{Example 2: Invalid Instrumental Variables}
%%%%%%%%%%%%%%%%%%%%%%%%%%%%%%%%%%%%%%%%%%%%%%%%%%%%%%%%%%%%%%%%%%
Suppose that there is endogeneity in the linear model $y_{i}=x_{i}\beta_{0} + \varepsilon_{i}$, where $y_{i},x_{i}\in\mathbb{R}$ and $Ex_{i}\varepsilon_{i}\neq 0$, so that the OLS estimator is inconsistent for $\beta_{0}$. Suppose that we have two instruments, $z_{1i}$ and $z_{2i}$. We can estimate $\beta_{0}$ using both instruments by GMM. The moment function is
\begin{equation}
g(X_{i},\beta) = \left(
                   \begin{array}{c}
                     z_{1i}(y_{i}-x_{i}\beta) \\
                     z_{2i}(y_{i}-x_{i}\beta) \\
                   \end{array}
                 \right),
\end{equation}
where $X_{i}=(y_{i},x_{i},z_{1i},z_{2i})'$. This moment function is correctly specified when both instruments are valid, i.e., $Ez_{1i}\varepsilon_{i}=Ez_{2i}\varepsilon_{i}=0$. In practice, a commonly used weight matrix is $W_{n}=(n^{-1}\sum_{i=1}^{n}\mathbf{z}_{i}\mathbf{z}_{i}')^{-1}$, where $\mathbf{z}_{i}=(z_{1i},z_{2i})'$. With this choice of the weight matrix, the one-step GMM estimator $\hat{\beta}_{(1)}$ is equivalent to the 2SLS estimator. If at least one of the instruments is invalid, then only the Hall-Inoue variance estimator $\hat{\Sigma}_{MR}$ is consistent for the true asymptotic variance of $\hat{\beta}_{(1)}$. Neither the conventional GMM variance estimator nor the 2SLS variance estimator is consistent.\footnote{Maasoumi and Phillips (1982) points out that the calculation of the asymptotic variance of overidentified and misspecified IV estimator is very complicated. Their asymptotic variance is a special case of Hall and Inoue (2003).}

Let the DGP be
\begin{eqnarray}
 && y_{i} = x_{i}\beta_{0} + \varepsilon_{i};\hspace{1em} x_{i} = z_{1i}\gamma_{1} + z_{2i}\gamma_{2} + u_{i},\hspace{1em}z_{2i} = z_{2i}^{0} + \frac{\delta}{(e-1)e}\varepsilon_{i};
\label{DGP2}
\\
\nonumber &&\varepsilon_{i} = \varepsilon_{i}^{0}-e^{0.5};\hspace{1em}u_{i}=u_{i}^{0}-e^{0.5};\\
\nonumber &&\left(
              \begin{array}{c}
                z_{1i} \\
                z_{2i}^{0} \\
              \end{array}
            \right)\sim N\left(\left(
                                                   \begin{array}{c}
                                                     0 \\
                                                     0 \\
                                                   \end{array}
                                                 \right)
,\left(
   \begin{array}{cc}
     1 & 0 \\
     0 & 1 \\
   \end{array}
 \right)\right),\hspace{1em}\left(
                \begin{array}{c}
                  \varepsilon_{i}^{0} \\
                  u_{i}^{0} \\
                \end{array}
              \right)\sim \log N\left(\left(
                                        \begin{array}{c}
                                          0 \\
                                          0 \\
                                        \end{array}
                                      \right),\left(
                                                \begin{array}{cc}
                                                  1 & .99\\
                                                  .99 & 1 \\
                                                \end{array}
                                              \right)\right),
\end{eqnarray}
where $(z_{1i},z_{2i}^{0})'$, $(\varepsilon_{i}^{0},u_{i}^{0})'$ are iid. The error terms are log-normally distributed with the mean zero. This DGP satisfies $Ex_{i}\varepsilon_{i}\neq0$, $Ez_{1i}\varepsilon_{i}=0$, and $Ez_{2i}\varepsilon_{i}=\delta$, where $\delta$ measures a degree of misspecification. Therefore, the instrument $z_{1i}$ is valid, while $z_{2i}$ may not. Let $\beta_{0}=0$ for simplicity. The probability limit of $\hat{\beta}_{(1)}$ is
\begin{equation}
\beta_{0(1)} = \frac{\left(\left(1+\frac{\delta^{2}}{(e-1)e}\right)\gamma_{2}+\frac{\delta}{(e-1)e}\rho_{\varepsilon u}\right)\delta}{\left(1+\frac{\delta^{2}}{(e-1)e}\right)\gamma_{1}^{2}+\left(\left(1+\frac{\delta^{2}}{(e-1)e}\right)\gamma_{2}+\frac{\delta}{(e-1)e}\rho_{\varepsilon u}\right)^{2}},
\label{beta0(1)}
\end{equation}
where $\rho_{\varepsilon u}=E\varepsilon_{i}u_{i}$. The pseudo-true value $\beta_{0(1)}$ depends on $\delta$, $\rho_{\varepsilon u}$, $\gamma_{1}$ and $\gamma_{2}$. Thus, it is different from $\beta_{0}=0$ in general. However, larger misspecification does not necessarily imply larger potential bias in the pseudo-true value. To see this, let
\begin{equation}
\gamma_{2} = -\frac{\delta\rho_{\varepsilon u}}{(e-1)e+\delta^{2}}.
\label{gamma2}
\end{equation}
Then $\beta_{0(1)}=\beta_{0}=0$ regardless of the value of $\delta$, $\rho_{\varepsilon u}$, and $\gamma_{1}$. Therefore, we can consistently estimate the structural parameter even with invalid instrument in this special case. Moreover, this particular choice of $\gamma_{2}$ can be considered as a strong but potentially invalid instrument. Let $\gamma_{1}=0.25$ so that the first instrument $z_{1i}$ is relatively weak.\footnote{The strength of instruments depends on the magnitude of the reduced form coefficient as well as the number of instruments, e.g., Hahn and Hausman (2002, 2005) and Guggenberger (2008). Since the weak instruments problem is not the main issue of this paper, I do not further investigate it.}  When $\delta=0$, then $\gamma_{2}=0$ so that $z_{2i}$ has no explanatory power. However, the instrument becomes stronger as $\delta$ deviates from zero given $\rho_{\varepsilon u}$ is not zero. We can significantly improve the finite sample coverage probability of CI's by using this instrument. Monte Carlo simulation results support this thought experiment.

Table \ref{table_ex2} shows the coverage probabilities of 90\% and 95\% CI's based on the one-step GMM estimator, $\hat{\beta}_{(1)}$ with $\gamma_{1}=0.25$ and $\gamma_{2}$ in \eqref{gamma2}. First, consider the case when $\delta=0$ so that both the instruments are valid but the second one has no explanatory power. The bootstrap CI's provide more accurate coverage than the asymptotic CI's when the model is correctly specified, but the bootstrap does not solve the problem of using a relatively weak instrument, see Hall and Horowitz (1996) for more discussions. Interestingly, the MR CI's show better performance than the conventional CI's when $n=50$ and $n=200$. This finding further supports the use of the MR CI's in practice, especially when one suspects an over-rejection of the $J$ test. There is a noticeable size distortion in the reported $J$ tests. The Hall-Horowitz bootstrap $J$ test shows smaller size distortion than the asymptotic one. Note that the MR bootstrap is not for the $J$ test, because it does not impose the correct specification of the model in implementing the bootstrap.

Now consider the misspecified cases, $\delta=0.25$ and $\delta=0.5$. By using the invalid but relatively strong instrument, the coverage of the MR CI's improves overall. $CI_{MR}^{*}$ performs better than $CI_{MR}$ regardless of misspecification, and there is a significant improvement even when $n=1,000$. In contrast, the conventional CI's are first-order invalid. The $J$ tests seem less powerful to reject the null hypothesis compared to Example 1 (Table \ref{table_ex1}). Furthermore, the Hall-Horowitz bootstrap $J$ test are less powerful than the asymptotic $J$ test.

\begin{figure}[p!]
	\centering
    \includegraphics[width=120mm]{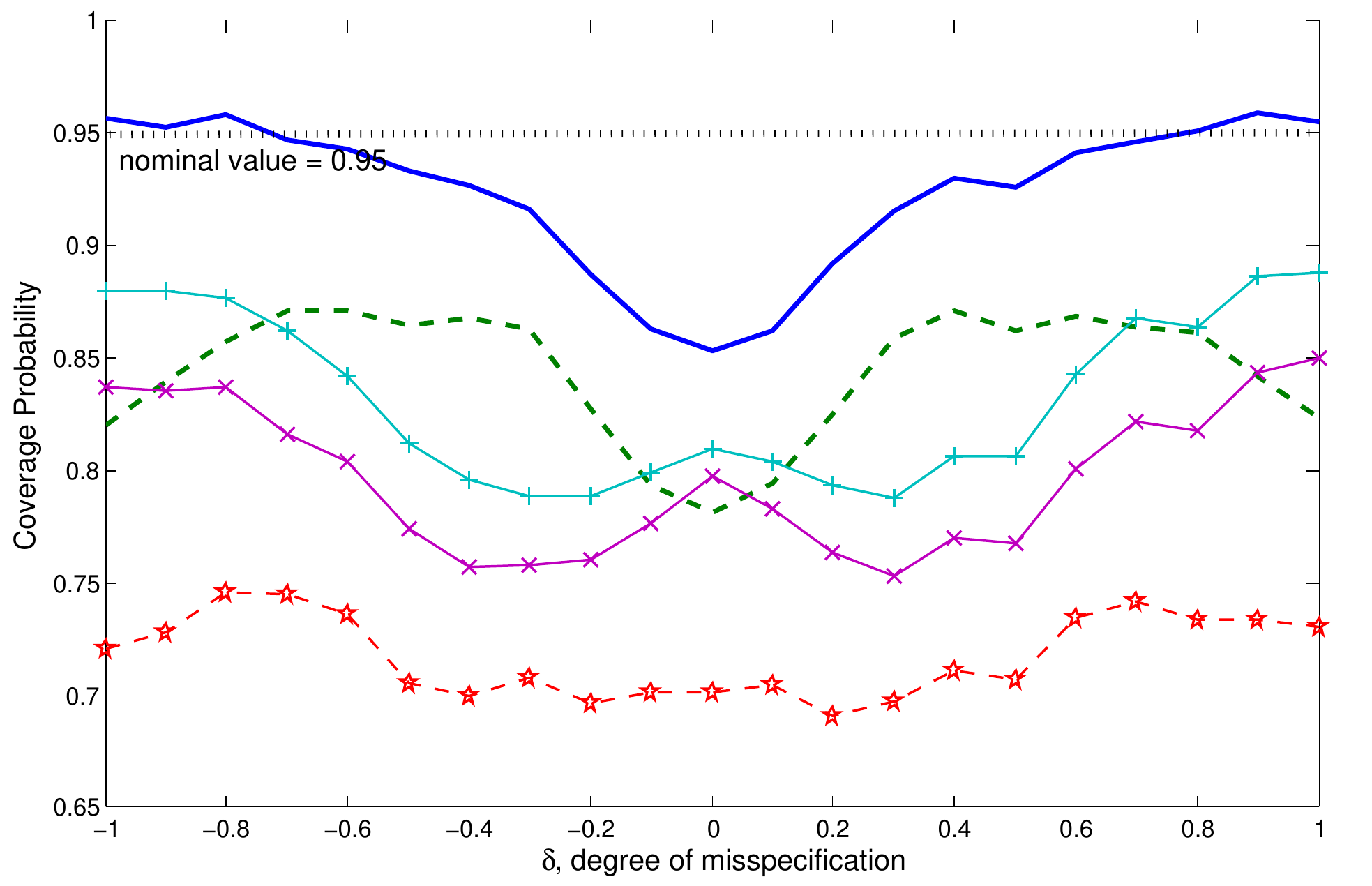}
    \caption{Coverage Probabilities of 95\% Confidence Intervals for $\beta_{0(1)}$ based on the One-step GMM Estimator, $\hat{\beta}_{(1)}$, $n=200$ in Example 2 DGP \eqref{DGP2}: $CI_{MR}^{*}$ (solid), $CI_{MR}$ (dashed), $CI_{C}$ (dashed with stars), $CI_{HH}^{*}$ (solid with +'s), $CI_{BN}^{*}$ (solid with x's)}
    \label{fig_ex2_D5}
\end{figure}

Figure \ref{fig_ex2_D5} shows the coverage probabilities of the CI's over different degrees of misspecification. It reinforces the previous finding: (i) The ability of achieving asymptotic refinements of the bootstrap CI's is clearly demonstrated at $\delta=0$, and $CI_{MR}^{*}$ maintains the ability regardless of misspecification, and (ii) the MR CI's may perform even better than the conventional CI's under correct specification.

%%%%%%%%%%%%%%%%%%%%%%%%%%%%%%%%%%%%%%%%%%%%%%%%%%%%%%%%
\subsection{Power}
%%%%%%%%%%%%%%%%%%%%%%%%%%%%%%%%%%%%%%%%%%%%%%%%%%%%%%%%

Asymptotic refinements of the bootstrap focus on the size, not the power of $t$ tests. Nevertheless, one may wonder the power property of the asymptotic and bootstrap $t$ tests. The null hypothesis is $H_{0}: \theta=\theta_{0(j)}$ for $j=1,2$. Similar to the CI's, we consider five types of two-sided symmetric $t$ tests. We have two $t$ statistics, $T_{C(j)}$ and $T_{MR(j)}$:
\begin{equation}
\nonumber T_{C(j)} = \frac{\hat{\theta}_{(j)}-\theta_{0(j)}}{\sqrt{\hat{\Sigma}_{C(j)}/n}}, \hspace{1em}  T_{MR(j)} = \frac{\hat{\theta}_{(j)}-\theta_{0(j)}}{\sqrt{\hat{\Sigma}_{MR(j)}/n}},
\end{equation}
where $\hat{\Sigma}_{C(j)}$ and $\hat{\Sigma}_{MR(j)}$ are the conventional variance estimator and the Hall-Inoue variance estimator, respectievly. Let the asymptotic significance level be $\alpha$. The conventional asymptotic $t$ test rejects the null if $|T_{C(j)}|>z_{\alpha/2}$, and is denoted by $t_{C}$. The MR asymptotic $t$ test rejects the null if $|T_{MR(j)}|>z_{\alpha/2}$, and is denoted by $t_{MR}$. The Hall-Horowitz and the Brown-Newey bootstrap $t$ tests reject the null if $|T_{C(j)}|>z_{|T_{HH}|,\alpha}^{*}$ and $|T_{C(j)}|>z_{|T_{BN}|,\alpha}^{*}$, and are denoted by $t_{HH}^{*}$ and $t_{BN}^{*}$, respectively. Finally, the MR bootstrap $t$ test rejects the null if $|T_{MR(j)}|>z_{|T_{MR}|,\alpha}$, and is denoted by $t_{MR}^{*}$.

\begin{figure}[p!]
  \centering
  \subfloat[Panel 1: $\delta=0$, $n=200$]{\label{fig_EX1_2sc_del0n200}\includegraphics[width=85mm]{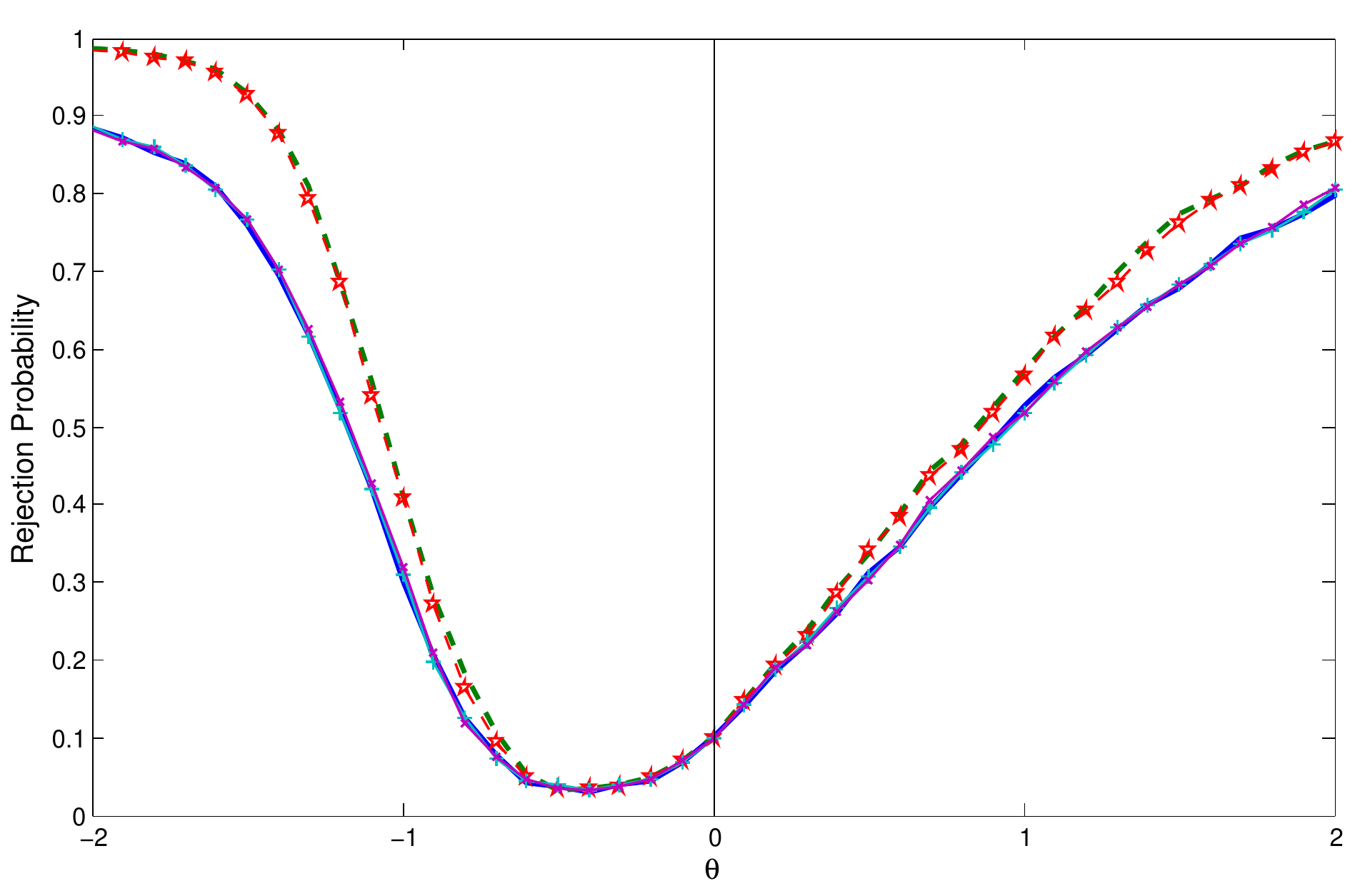}}\\
  \subfloat[Panel 2: $\delta=0.6$, $n=200$]{\label{fig_EX1_2sc_del06n200}\includegraphics[width=85mm]{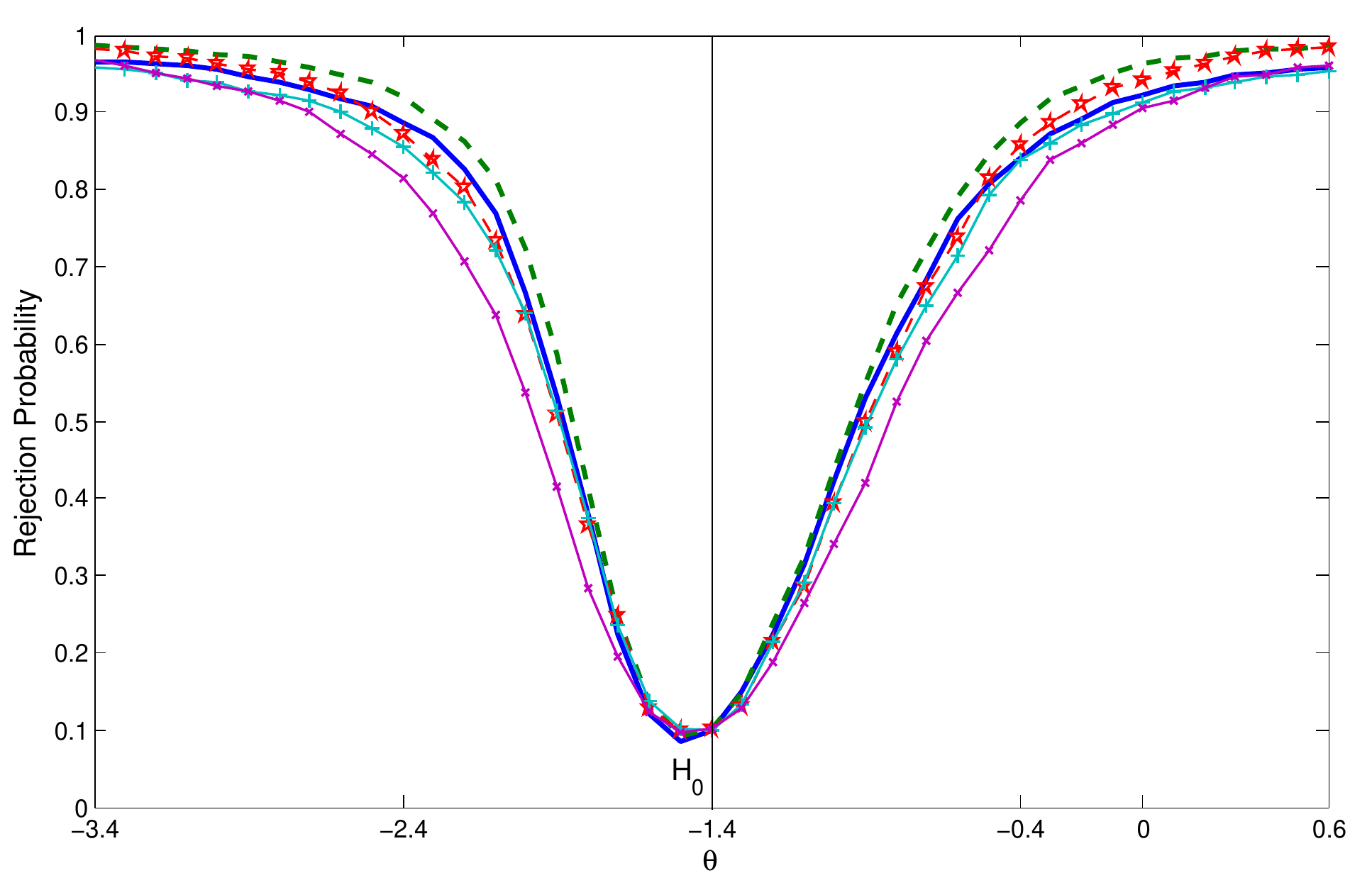}}\\
  \subfloat[Panel 3: $\delta=-0.6$, $n=200$]{\label{fig_EX1_2sc_delN06n200}\includegraphics[width=85mm]{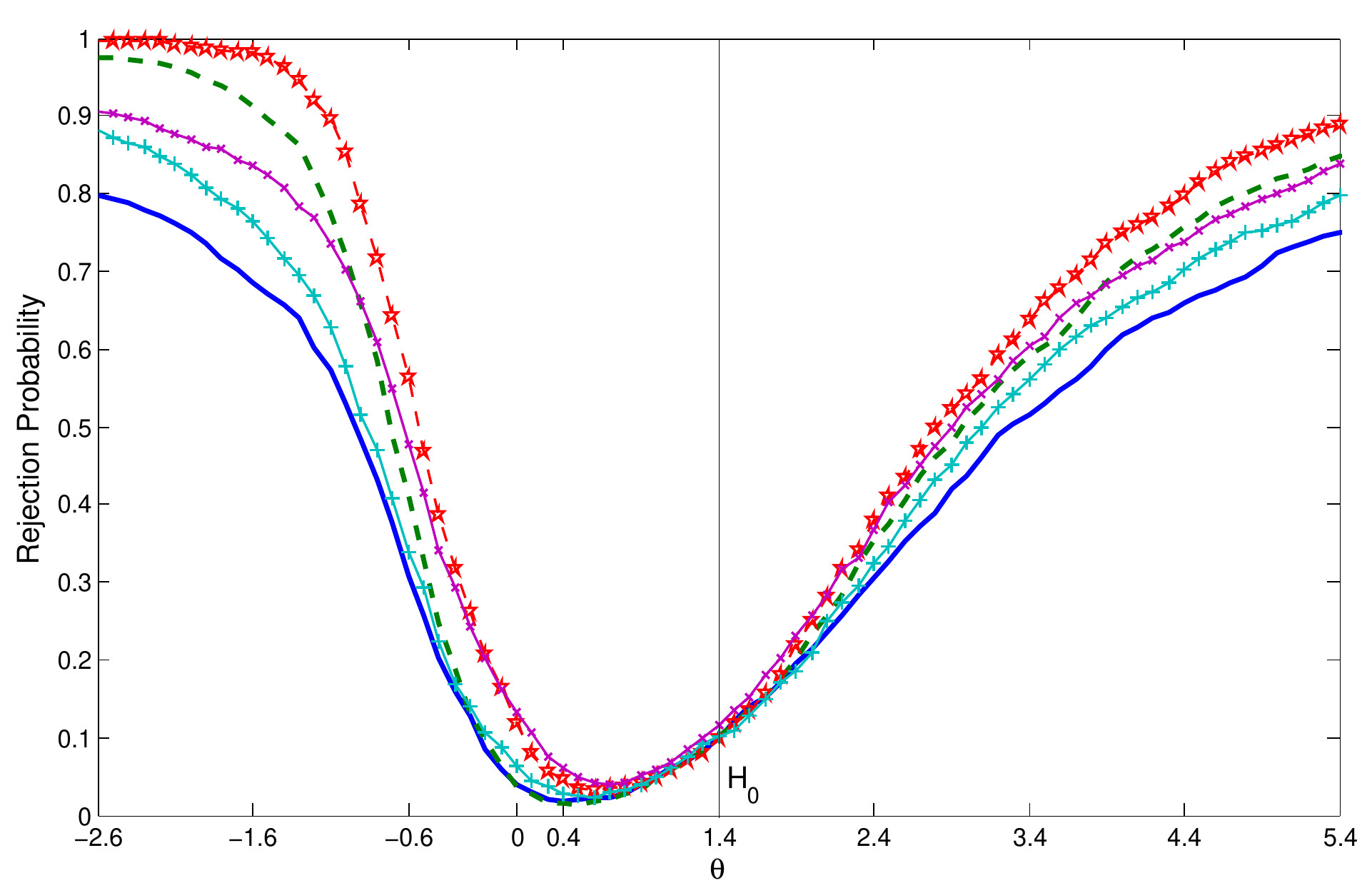}}\\
  \caption{(Size corrected) Power curves of $t$ statistics that test $H_{0}: \theta=\theta_{0(2)}$ with 10\% asymptotic significance level in Example 1 with $\delta=0,0.6,-0.6$ and $n=200$: $t_{MR}^{*}$ (solid), $t_{MR}$ (dashed), $t_{C}$ (dashed with stars), $t_{HH}^{*}$ (solid with +'s), $t_{BN}^{*}$ (solid with x's)}
  \label{fig_pw1}
\end{figure}

\begin{figure}[p!]
  \centering
  \subfloat[Panel 1: $\delta=0$, $n=200$]{\label{fig_EX2_5sc_del0n200}\includegraphics[width=85mm]{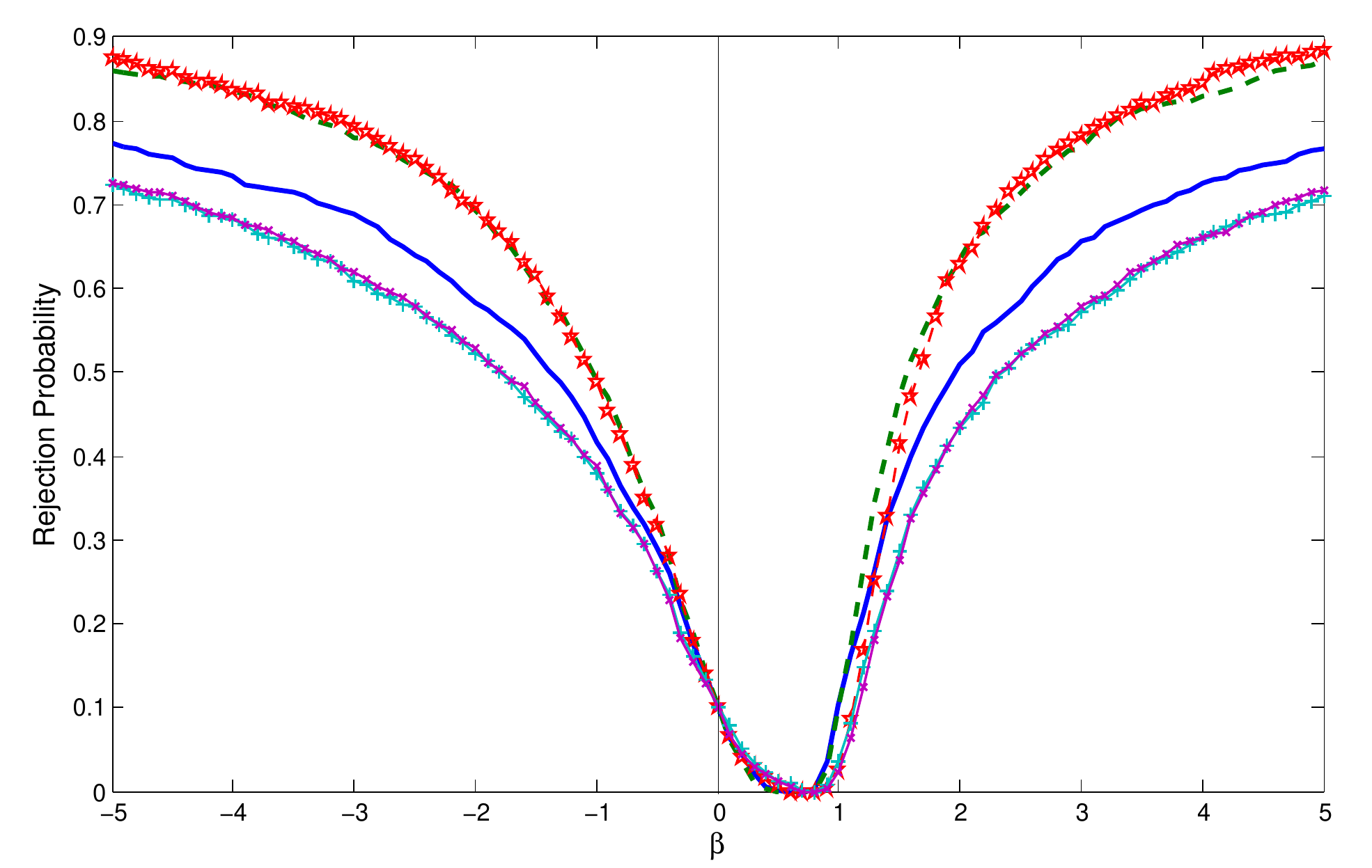}}\\
  \subfloat[Panel 2: $\delta=0.25$, $n=200$]{\label{fig_EX2_5sc_del025n200}\includegraphics[width=85mm]{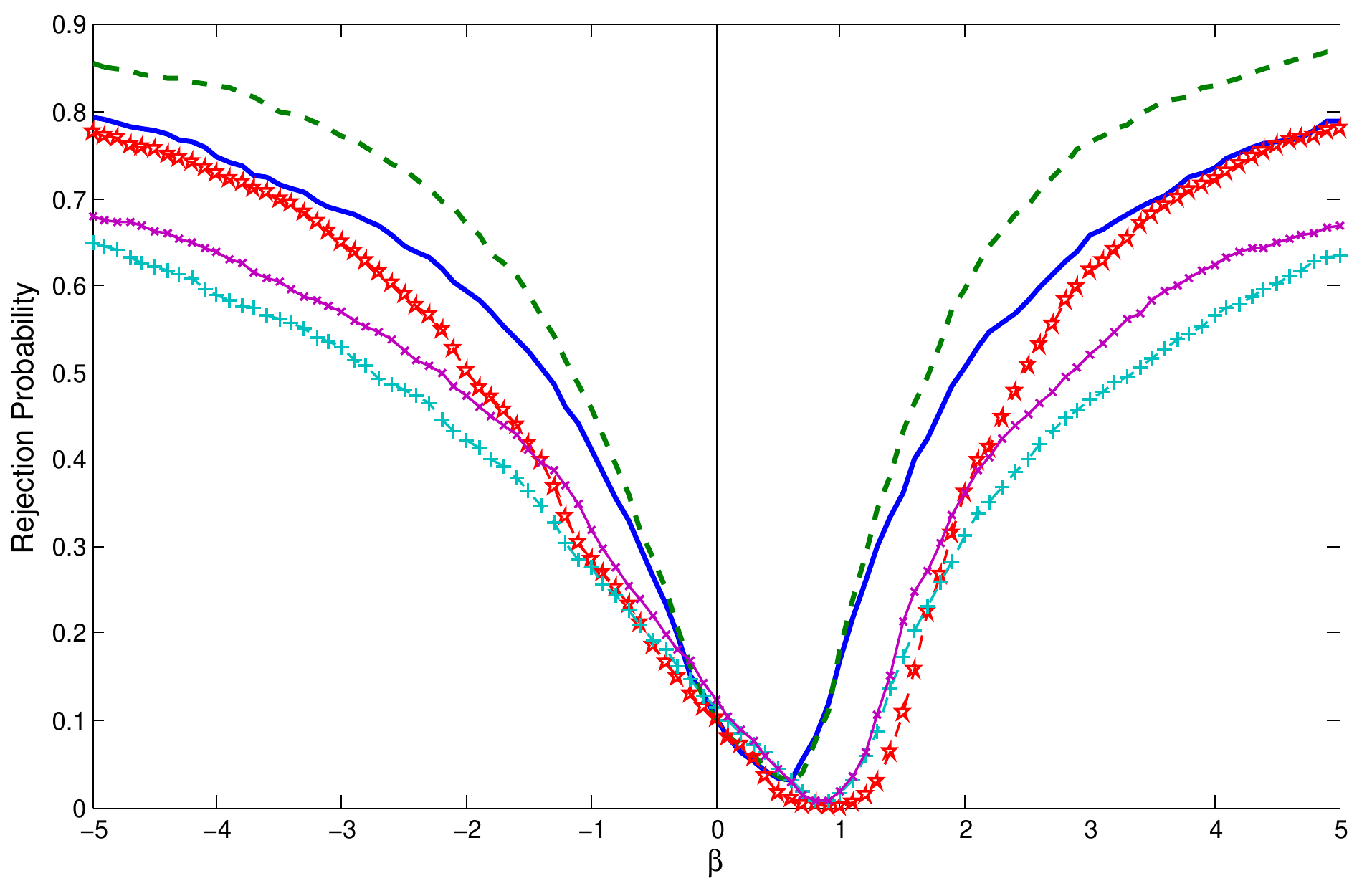}}\\
  \subfloat[Panel 3: $\delta=0.5$, $n=200$]{\label{fig_EX2_5sc_del05n200}\includegraphics[width=85mm]{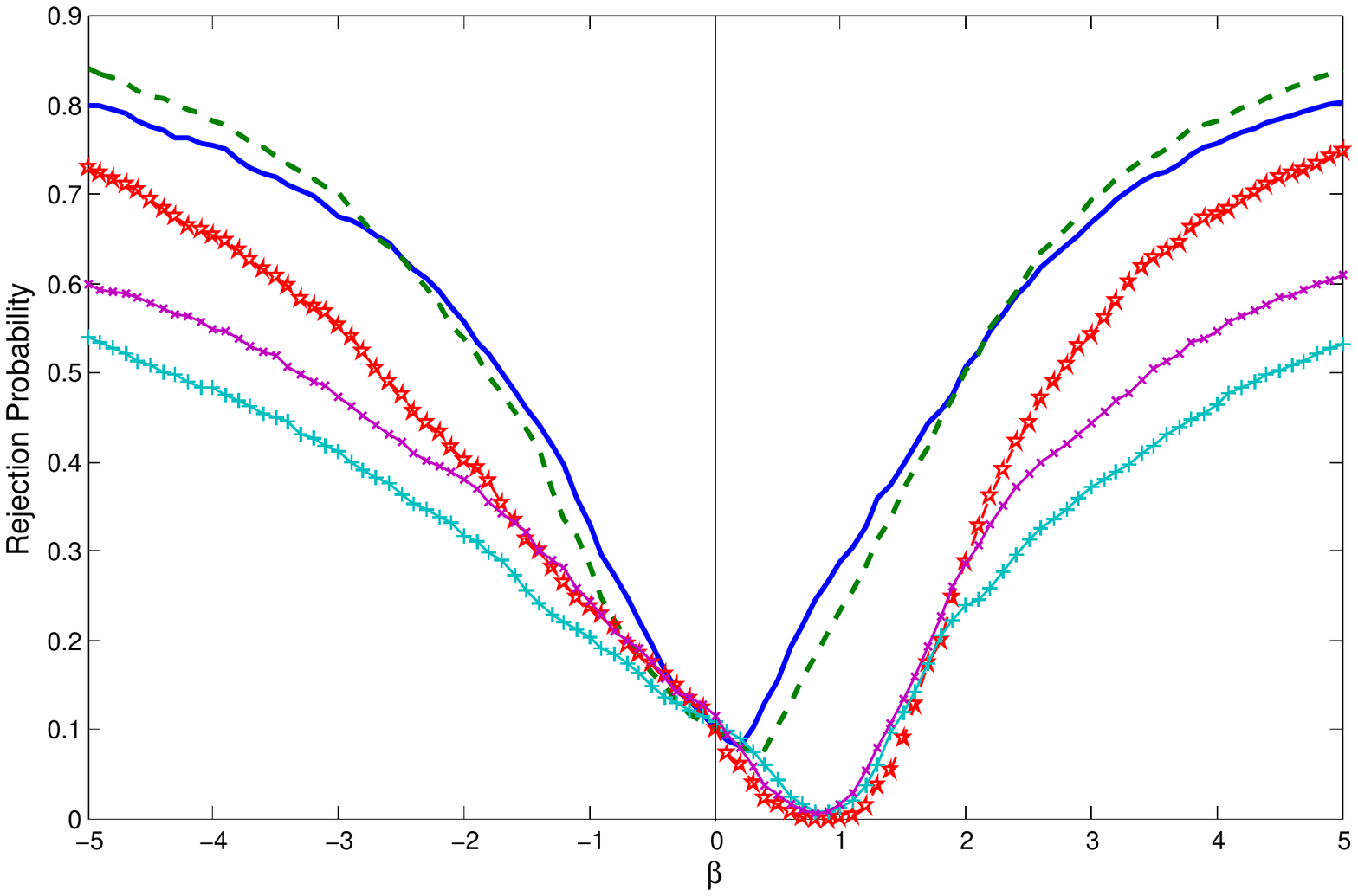}}\\
  \caption{(Size corrected) Power curves of $t$ statistics that test $H_{0}: \beta=0$ with 10\% asymptotic significance level in Example 2 DGP 2 with $\delta=0,0.25,0.5$ and $n=200$: $t_{MR}^{*}$ (solid), $t_{MR}$ (dashed), $t_{C}$ (dashed with stars), $t_{HH}^{*}$ (solid with +'s), $t_{BN}^{*}$ (solid with x's)}
  \label{fig_pw2}
\end{figure}

Figures \ref{fig_pw1} and \ref{fig_pw2} show the power curves of the $t$ statistics in Examples 1 and 2. Since the $t$ tests show large size distortion as we saw in the previous section, I use the 10\% size-corrected critical values for the asymptotic and bootstrap $t$ tests. The number of Monte Carlo repetition (r) is 1,000 and the number of bootstrap replication (B) is 1,000. For each generated sample, the $t$ statistics are evaluated at various values of $\theta$ around the null and the rejection frequency of the $t$ tests is computed using the size-corrected critical values.

The conclusion is mixed. We find from the figures that under correct specification, (i) the asymptotic $t$ tests show better power properties than the bootstrap $t$ tests ($t_{MR}$ over $t_{MR}^{*}$; $t_{C}$ over $t_{HH}^{*}$ and $t_{BN}^{*}$), but (ii) it is difficult to rank between the asymptotic $t$ tests ($t_{MR}$ and $t_{C}$), and among the bootstrap $t$ tests ($t_{MR}^{*}$, $t_{HH}^{*}$, and $t_{BN}^{*}$). Under misspecification, the conventional asymptotic and bootstrap $t$ tests are inconsistent. The power of the MR asymptotic and bootstrap $t$ tests are not necessarily weaker than the ones using standard $t$ statistic (Figure \ref{fig_pw2} Panels 2 and 3). In addition, the MR bootstrap $t$ test can be more powerful than the MR asymptotic $t$ test (Figure 4 Panel 3).

%%%%%%%%%%%%%%%%%%%%%%%%%%%%%%%%%%%%%%%%%%%%%%%%%%%%%%%%
\section{Conclusion}
\label{Conc}
%%%%%%%%%%%%%%%%%%%%%%%%%%%%%%%%%%%%%%%%%%%%%%%%%%%%%%%%
Bootstrap critical values allow more accurate inferences and CI's than the asymptotic critical values. To get the bootstrap refinements for GMM estimators, an ad hoc procedure called recentering has been considered as critical in the existing literature. In addition, the conventional bootstrap methods are not robust to unknown model misspecification. In contrast, the proposed MR bootstrap achieves the same rate of asymptotic refinements without recentering, and without assuming correct specification of the model. The key idea is to link the misspecified moment condition in the bootstrap world to the large sample theory of GMM under misspecification of Hall and Inoue (2003).

Possible extensions of this paper would be (i) to see whether the MR bootstrap still works conditional on the event that the $J$ test fails to reject the null as this is likely to happen in practice, and (ii) to apply the MR bootstrap to the GEL estimators.

%%%%%%%%%%%%%%%%%%%%%%%%%%%%%%%%%%%%%%%%%%%%%%%%%%%%%%%%
\section*{Acknowledgment}
%%%%%%%%%%%%%%%%%%%%%%%%%%%%%%%%%%%%%%%%%%%%%%%%%%%%%%%%
I am very grateful to Bruce Hansen and Jack Porter for their guidance and helpful comments. I also thank Ken West, Xiaoxia Shi, Don Andrews, Ping Yu, and James Morley, as well as seminar participants at Auckland, Iowa, Sogang, Sungkyunkwan, Sydney, UNSW, Wisconsin-Madison, and Yale for their discussions and suggestions. An earlier version of this paper was presented at the 2011 NASM, 2011 AMES, and 2011 Midwest Econometrics Group. Finally, I thank the co-editor, the associate editor, and three referees for their comments and suggestions that greatly improved the presentation of the paper.

\appendix
%%%%%%%%%%%%%%%%%%%%%%%%%%%%%%%%%%%%%%%%%%%%%%%%%%%%%%%%
\section{Appendix: Lemmas and Proofs}
%%%%%%%%%%%%%%%%%%%%%%%%%%%%%%%%%%%%%%%%%%%%%%%%%%%%%%%%

The proofs of the Theorem and Lemmas are analogous to those of Hall and Horowitz (1996) and Andrews (2002) by allowing possible model misspecification. Throughout the Appendix, write $g_{i}(\theta)=g(X_{i},\theta)$, $g_{i}^{*}(\theta)=g(X_{i}^{*},\theta)$, $G_{i}(\theta)=G(X_{i},\theta)$, $G_{i}^{*}(\theta)=G(X_{i}^{*},\theta)$, $f_{i}(\theta)=f(X_{i},\theta)$, and $f_{i}^{*}(\theta)=f(X_{i}^{*},\theta)$ for notational brevity.

\subsection{Proof of Theorem 1}

The usage of the Hall-Inoue variance estimators in constructing the sample and bootstrap versions of the $t$ statistic without recentering the bootstrap moment function is taken into account by Lemmas \ref{L10} and \ref{L12}. Once we establish the Edgeworth expansions of $T_{MR(j)}$ and $T_{MR(j)}^{*}$ for $j=1,2$, the proof of the Theorem is the same with that of Theorem 2(c) of Andrews (2002) with his Lemmas 13 and 16 replaced by our Lemmas \ref{L10} and \ref{L12}. His proof relies on the argument of Hall (1988, 1992)'s methods developed for ``smooth functions of sample averages,'' for iid data. \qed

\subsection{Lemmas}

Lemma \ref{L0} modifies Lemmas 1, 2, 6, and 7 of Andrews (2002) for nonparametric iid bootstrap under possible misspecification. The modified Lemmas 1, 2, 6, and 7 are denoted by AL1, AL2, AL6, and AL7, respectively. In addition, Lemma 5 of Andrews (2002) is denoted by AL5 without modification. The complete proofs of the Lemmas are in a separate supplementary appendix available at the author's website: \href{sites.google.com/site/misspecified/}{sites.google.com/site/misspecified/}

\begin{lemma}\
  \begin{description}
    \item[(a)] Lemma 1 of Andrews (2002) holds by replacing $\widetilde{X}_{i}$ and $N$ with $X_{i}$ and $n$, respectively, under our Assumption 1.
    \item[(b)] Lemma 2 of Andrews (2002) for $j=1$ holds under our Assumptions 1-3.
    \item[(c)] Lemma 6 of Andrews (2002) holds by replacing $\widetilde{X}_{i}$ and $N$ with $X_{i}$ and $n$, respectively, and by letting $l=1$ and $\gamma=0$, under our Assumption 1.
    \item[(d)] Lemma 7 of Andrews (2002) for $j=1$ holds by replacing $\widetilde{X}_{i}$ and $N$ with $X_{i}$ and $n$, respectively, and by letting $l=1$ and $\gamma=0$, under our Assumptions 1-3.
  \end{description}
\label{L0}
\end{lemma}
\vspace{-1em}
\noindent
\textit{Proof.}
(a) Assumption 1 of Andrews (2002) is satisfied if our Assumption 1 holds. Thus, Lemma 1 of Andrews (2002) holds.\\
(b) We use the proof of Lemma 2 of Andrews (2002) which relies on that of Lemma 2 of Hall and Horowitz (1996). Since their proof does not require $Eg(X_{i},\theta_{0})=0$, the Lemma holds under our Assumptions 1-3.\\
(c) Assumption 1 of Andrews (2002) is satisfied if our Assumption 1 holds. Thus, Lemma 6 of Andrews (2002) holds for the nonparametric iid bootstrap.\\
(d) We use the proof of Lemma 7 of Andrews (2002) which relies on that of Lemma 8 of Hall and Horowitz (1996). Since their proof does not require $Eg(X_{i},\theta_{0})=0$, the Lemma holds for the nonparametric iid bootstrap under our Assumptions 1-3. \qed

Lemmas \ref{L3}-\ref{L4} prove that the one-step and two-step GMM estimators are consistent for the (pseudo-)true values, $\theta_{0(1)}$ and $\theta_{0(2)}$, respectively, under possible misspecification.

%%%%%%%%%%%%%%%%%%%%%%%%%%%%%%%%%%%%%%%%%%%%%%%%%%%%%
\begin{lemma}
Suppose Assumptions 1-3 hold. Then, for all $c\in[0,1/2)$ and all $a\geq 0$,
\[\lim_{n\rightarrow\infty}n^{a}P(\|\hat{\theta}_{(1)}-\theta_{0(1)}\|>n^{-c})=0.\]
\label{L3}
\end{lemma}
\vspace{-3em}
\noindent
\textit{Proof.}
The proof is similar to that of Lemma 3 of Andrews (2002) with the following exceptions. Instead of his (9.25), we have
\begin{equation}
\hat{\theta}_{(1)}-\theta_{0(1)}=-\left(\frac{\partial^{2}}{\partial\theta\partial\theta'}J_{n}(\tilde{\theta})\right)^{-1}\frac{\partial}{\partial\theta}J_{n}(\theta_{0(1)})
\end{equation}
with probability $1-o(n^{-a})$, where
\begin{eqnarray}
\frac{\partial}{\partial\theta}J_{n}(\theta_{0(1)})&=&\left\{G_{0(1)}'(g_{n}(\theta_{0(1)})-g_{0(1)})+(G_{n}(\theta_{0(1)})-G_{0(1)})'g_{n}(\theta_{0(1)})\right\},\\
\nonumber \frac{\partial^{2}}{\partial\theta\partial\theta'}J_{n}(\theta)&\equiv&2\tilde{H}_{n}(\theta,I_{L_{g}})=2\left\{\left(g_{n}(\theta)'\otimes I_{L_{\theta}}\right)G_{n}^{(2)}(\theta)+G_{n}(\theta)'G_{n}(\theta)\right\},
\end{eqnarray}
and $\tilde{\theta}$ is between $\hat{\theta}_{(1)}$ and $\theta_{0(1)}$ and may differ across rows. Note that the first and second derivatives of $J_{n}(\theta)$ include additional terms that do not appear under correct specification, $g_{0(1)}=0$. Then, instead of his (9.26), we have
\begin{eqnarray}
&&\lim_{n\rightarrow\infty} n^{a}P\left(\left\|\tilde{H}_{n}(\tilde{\theta},I_{L_{g}})-\tilde{H}_{n}(\theta_{0(1)},I_{L_{g}})\right\|>\varepsilon\right)=0,
\label{e1}\\
\nonumber &&\lim_{n\rightarrow\infty} n^{a}P\left(\left\|\tilde{H}_{n}(\theta_{0(1)},I_{L_{g}})-H_{0(1)}\right\|>\varepsilon\right)=0,
\label{e2}\\
\nonumber &&\lim_{n\rightarrow\infty}
n^{a}P\left(\left\|G_{n}(\theta_{0(1)})-G_{0(1)}\right\|>n^{-c}\right)=0,
\label{e3}\\
\nonumber &&\lim_{n\rightarrow\infty}
n^{a}P\left(\left\|g_{n}(\theta_{0(1)})-g_{0(1)}\right\|>n^{-c}\right)=0.
\label{e4}
\end{eqnarray}
\eqref{e1} can be shown by applying the triangle and Cauchy-Schwarz inequalities as well as AL1.\qed

%%%%%%%%%%%%%%%%%%%%%%%%%%%%%%%%%%%%%%%%%%%%%%%%%%%%%%%
\begin{lemma}
Suppose Assumptions 1-3 hold. Then, for all $c\in[0,1/2)$ and all $a\geq 0$,
\[\lim_{n\rightarrow\infty}n^{a}P(\|\hat{\theta}_{(2)}-\theta_{0(2)}\|>n^{-c})=0.\]
\label{L4}
\end{lemma}
\vspace{-3em}
\noindent
\textit{Proof.}
The proof is similar to that of Lemma 4 of Andrews (2002), except that we apply AL1 and Lemma \ref{L3} instead of his Lemma 1 and Lemma 3.  \qed

%\vspace{-2em}
Lemmas \ref{L8}-\ref{L9} are the bootstrap versions of Lemmas \ref{L3}-\ref{L4}, respectively, and consistency of the MR bootstrap is established under possible misspecification. Note that the bootstrap GMM estimators are different from the Hall-Horowitz bootstrap GMM estimators, which use the recentered bootstrap moment function.

%%%%%%%%%%%%%%%%%%%%%%%%%%%%%%%%%%%%%%%%%%%%%%%%%%%%%%%
\begin{lemma}
Suppose Assumptions 1-3 hold. Then, for all $c\in[0,1/2)$ and all $a\geq 0$,
\[\lim_{n\rightarrow\infty}n^{a}P(P^{*}(\|\hat{\theta}^{*}_{(1)}-\hat{\theta}_{(1)}\|>n^{-c})>n^{-a})=0.\]
\label{L8}
\end{lemma}
\vspace{-3em}
\noindent
\textit{Proof.}
First, we prove the result with $n^{-c}$ replaced by a fixed $\varepsilon>0$. The proof is similar to that of Lemma 9 of Andrews (2002) except that we use AL2 and AL7 instead of his Lemma 2 and Lemma 7.

Next, we prove the result stated in the Lemma. Write $J(\theta)\equiv J(\theta,I_{L_{g}})$ and $J^{*}_{n}(\theta)\equiv J^{*}_{n}(\theta,I_{L_{g}})$ for notational brevity. The first-order condition is $(\partial/\partial\theta) J_{n}^{*}(\hat{\theta}^{*}_{(1)})=G_{n}^{*}(\hat{\theta}^{*}_{(1)})'g_{n}^{*}(\hat{\theta}^{*}_{(1)})=0$ with $P^{*}$ probability $1-o(n^{-a})$ except, possibly, if $\chi$ is in a set of $P$ probability $o(n^{-a})$. By the mean value theorem,
\begin{eqnarray}
\hat{\theta}^{*}_{(1)}-\hat{\theta}_{(1)}&=&-\left(\frac{\partial^{2}}{\partial\theta\partial\theta'}J_{n}^{*}(\tilde{\theta}^{*})\right)^{-1}\frac{\partial}{\partial\theta}J_{n}^{*}(\hat{\theta}_{(1)}),
\end{eqnarray}
with $P^{*}$ probability $1-o(n^{-a})$ except, possibly, if $\chi$ is in a set of $P$ probability $o(n^{-a})$, where $\tilde{\theta}^{*}$ is between $\hat{\theta}^{*}_{(1)}$ and $\hat{\theta}_{(1)}$ and may differ across rows. Now the Lemma follows combining the following results:
\begin{eqnarray}
&&\lim_{n\rightarrow\infty}n^{a}P\left(P^{*}\left(\left\|\frac{\partial}{\partial\theta}J_{n}^{*}(\hat{\theta}_{(1)})\right\|>n^{-c}\right)>n^{-a}\right)=0,\\
\nonumber &&\lim_{n\rightarrow\infty}n^{a}P\left(P^{*}\left(\left\|\tilde{H}_{n}^{*}(\tilde{\theta}^{*},I_{L_{g}})-\tilde{H}_{n}^{*}(\theta_{0(1)},I_{L_{g}})\right\|>\varepsilon\right)>n^{-a}\right)=0,
\label{9231}\\
\nonumber &&\lim_{n\rightarrow\infty}n^{a}P\left(P^{*}\left(\left\|\tilde{H}_{n}^{*}(\theta_{0(1)},I_{L_{g}})-H_{0(1)}\right\|>\varepsilon\right)>n^{-a}\right)=0,
\label{9232}
\end{eqnarray}
where $\tilde{H}_{n}^{*}(\theta,I_{L_{g}})=(g_{n}^{*}(\theta)'\otimes I_{L_{\theta}})G_{n}^{(2)*}(\theta)+G_{n}^{*}(\theta)'G_{n}^{*}(\theta)$ and $(\partial^{2}/\partial\theta\partial\theta')J_{n}^{*}(\theta)=2\tilde{H}_{n}^{*}(\theta,I_{L_{g}})$. The proof follows that of Lemma \ref{L3} with some modifications for the bootstrap version using AL6. \qed

%%%%%%%%%%%%%%%%%%%%%%%%%%%%%%%%%%%%%%%%%%%%%%%%%%%%%%%%%
\begin{lemma}
Suppose Assumptions 1-3 hold. Then, for all $c\in[0,1/2)$ and all $a\geq 0$,
\[\lim_{n\rightarrow\infty}n^{a}P(P^{*}(\|\hat{\theta}_{(2)}^{*}-\hat{\theta}_{(2)}\|>n^{-c})>n^{-a})=0.\]
\label{L9}
\end{lemma}
\vspace{-3em}
\noindent
\textit{Proof.}
We first show that
\begin{equation}
\lim_{n\rightarrow\infty}n^{a}P(P^{*}(\|W^{*}_{n}(\hat{\theta}^{*}_{(1)})-W\|>n^{-c})>n^{-a})=0.
\label{l9e1}
\end{equation}
The proof is analogous to that of Lemma 4 in Andrews (2002), except that we use our Lemma \ref{L8} and AL6 instead of his Lemma 3 and Lemma 1, respectively. The rest of the proof is analogous to that of Lemma \ref{L8}. \qed

We now introduce some additional notation. Let $S_{n}$ be the vector containing the unique components of $n^{-1}\sum_{i=1}^{n}\left(f_{i}(\theta_{0(1)})',f_{i}(\theta_{0(2)})'\right)'$ on the support of $X_{i}$, and $S=ES_{n}$. Similarly, let $S^{*}_{n}$ denote the vector containing the unique components of $n^{-1}\sum_{i=1}^{n}\left(f_{i}^{*}(\hat{\theta}_{(1)})',f_{i}^{*}(\hat{\theta}_{(2)})'\right)'$ on the support of $X_{i}$, and $S^{*}=E^{*}S_{n}^{*}$. Note that the definitions of $S_{n}$ and $S^{*}_{n}$ are different from those of Hall and Horowitz (1996) and Andrews (2002), because they do not distinguish $\theta_{0(1)}$ and $\theta_{0(2)}$ by assuming the unique true value $\theta_{0}$. Under misspecification, $\theta_{0(1)}$ and $\theta_{0(2)}$ are different and thus, $\hat{\theta}_{(1)}$ and $\hat{\theta}_{(2)}$ have different probability limits. In addition, Hall and Horowitz (1996) and Andrews (2002) define $S_{n}^{*}$ by using the recentered moment function.

%%%%%%%%%%%%%%%%%%%%%%%%%%%%%%%%%%%%%%%%%%%%%%%%%%%%%%
\begin{lemma}
Let $\Delta_{n}$ and $\Delta_{n}^{*}$ denote $n^{1/2}(\hat{\theta}_{(j)}-\theta_{0(j)})$ and $n^{1/2}(\hat{\theta}^{*}_{(j)}-\hat{\theta}_{(j)})$, or $T_{MR(j)}$ and $T_{MR(j)}^{*}$ for $j=1,2$. For each definition of $\Delta_{n}$ and $\Delta_{n}^{*}$, there is an infinitely differentiable function $A(\cdot)$ with $A(S)=0$ and $A(S^{*})=0$ such that the following results hold.
\begin{description}
  \item[(a)] Suppose Assumptions 1-4 hold with $d_{1}\geq2a+2$, where $2a$ is some nonnegative integer. Then,
  \[\lim_{n\rightarrow\infty}\sup_{z}n^{a}|P(\Delta_{n}\leq z)-P(n^{1/2}A(S_{n})\leq z)|=0.\]
  \item[(b)] Suppose Assumptions 1-4 hold with $d_{1}\geq2a+2$, where $2a$ is some nonnegative integer. Then,
  \[\lim_{n\rightarrow\infty}n^{a}P\left(\sup_{z}|P^{*}(\Delta_{n}^{*}\leq z)-P^{*}(n^{1/2}A(S_{n}^{*})\leq z)|>n^{-a}\right)=0.\]
\end{description}
\label{L10}
\end{lemma}
\vspace{-1em}
\noindent
\textit{Proof.}
(a) The proof is analogous to that of Lemma 13 of Andrews (2002) which uses that of Proposition 1 of Hall and Horowitz (1996), except that it allows different probability limits for the one-step and the two-step GMM estimators. First, we show that $\hat{\theta}_{(j)}-\theta_{0(j)}$ can be approximated by a function of sample moments for $j=1,2$. We take the Taylor expansion of the first-order conditions up to order $d_{1}$. The proof for the one-step GMM is similar to that of Proposition 1 of Hall and Horowitz (1996). For the two-step GMM, write $J_{n}(\hat{\theta},\tilde{\theta})\equiv J_{n}(\hat{\theta},W_{n}(\tilde{\theta}))$ and let $(\partial_{1}/\partial\theta)J(\cdot,\cdot)$ denote the gradient of $J_{n}(\cdot,\cdot)$ with respect to its first argument. Then, $\partial_{1} J_{n}(\hat{\theta}_{(2)},\hat{\theta}_{(1)})/\partial\theta=0$ with probability $1-o(n^{-a})$ by the first-order condition of the two-step GMM. We take the Taylor expansion of $\partial_{1} J_{n}(\hat{\theta}_{(2)},\hat{\theta}_{(1)})/\partial\theta$ through order $d_{1}$ about $(\theta,\tilde{\theta})=(\theta_{0(2)},\theta_{0(1)})$, while Hall and Horowitz (1996) takes the Taylor expansion around $(\theta_{a},\theta_{b})=(\theta_{0},\theta_{0})$, the unique true value because $\theta_{0(2)}=\theta_{0(1)}$ under correct specification.

Andrews (2002) and Hall and Horowitz (1996) consider $T_{C(j)}$ while we consider $T_{MR(j)}$, but the proofs are similar because (i) the only difference is that the variance estimators are different, and (ii) the covariance matrix estimator, $\hat{\Sigma}_{MR(j)}$, is a function of $\hat{\theta}_{(j)}$, $j=1,2$, by construction. To ensure the existence of the derivatives of $T_{MR(j)}$, we need at least $d_{1}+1$ times differentiability of $g_{i}(\theta)$ with respect to $\theta$ because $\Sigma_{MR(j)}$ involves second derivatives of the moment function. By Assumption 3(c), this is satisfied.

(b) The proof for $\Delta_{n}^{*}=n^{1/2}(\hat{\theta}_{(j)}^{*}-\hat{\theta}_{(j)})$ for $j=1,2$, mimics that of Proposition 2 of Hall and Horowitz (1996) except that we take the Taylor expansion up to order $d_{1}$ rather than order 4. For the rest of the proof, observe that $\Delta_{n}^{*}$ has the same form of $\Delta_{n}$ by replacing $S_{n}$ and $\theta_{0(j)}$ with $S_{n}^{*}$ and $\hat{\theta}_{(j)}$, respectively, because $\Delta_{n}^{*}$ does not involve any recentering procedure as in Hall and Horowitz (1996). Therefore, the remainder of the proof proceeds as in the previous proof for part (a) of the Lemma. We use Lemmas \ref{L8}-\ref{L9} instead of Lemmas \ref{L3}-\ref{L4}. \qed

We define the components of the Edgeworth expansions of the test statistic $T_{MR(j)}$ and its bootstrap analog $T_{MR(j)}^{*}$. Let $\Psi_{n}=n^{1/2}(S_{n}-S)$ and $\Psi_{n}^{*}=n^{1/2}(S_{n}^{*}-S^{*})$. Let $\Psi_{n,k}$ and $\Psi_{n,k}^{*}$ denote the $k$th elements of $\Psi_{n}$ and $\Psi_{n}^{*}$, respectively. Let $\nu_{n,a}$ and $\nu_{n,a}^{*}$ denote vectors of moments of the form $n^{\alpha(m)}E\prod_{\mu=1}^{m}\Psi_{n,k_{\mu}}$ and $n^{\alpha(m)}E^{*}\prod_{\mu=1}^{m}\Psi^{*}_{n,k_{\mu}}$, respectively, where $2\leq m\leq 2a+2$, $\alpha(m)=0$ if $m$ is even, and $\alpha(m)=1/2$ if $m$ is odd. Let $\nu_{a}=\lim_{n\rightarrow\infty}\nu_{n,a}$. The limit exists under Assumption 1 of Andrews (2002), and thus under our Assumption 1.

Let $\pi_{i}(\delta,\nu_{a})$ be a polynomial in $\delta=\partial/\partial z$ whose coefficients are polynomials in the elements of $\nu_{a}$ and for which $\pi_{i}(\delta,\nu_{a})\Phi(z)$ is an even function of $z$ when $i$ is odd and is an odd function of $z$ when $i$ is even for $i=1,...,2a$, where $2a$ is an integer. The Edgeworth expansions of $T_{MR(j)}$ and $T_{MR(j)}^{*}$ depend on $\pi_{i}(\delta,\nu_{a})$ and $\pi_{i}(\delta,\nu_{n,a}^{*})$, respectively.

The following Lemma shows that the bootstrap moments $\nu_{n,a}^{*}$ are close to the population moments $\nu_{a}$ in large samples. The Lemma is an iid version of Lemma 14 of Andrews (2002).

%%%%%%%%%%%%%%%%%%%%%%%%%%%%%%%%%%%%%%%%%%%%%%%%%%%%%
\begin{lemma}
Suppose Assumptions 1 and 3 hold with $d_{2}\geq 2a+1$ for some $a\geq0$. Then, for all $c\in[0,1/2)$,
\[\lim_{n\rightarrow\infty}n^{a}P(\|\nu_{n,a}^{*}-\nu_{a}\|>n^{-c})=0.\]
\label{L11}
\end{lemma}
\vspace{-3em}
\noindent
\textit{Proof.}
Since $X_{i}$'s are iid by Assumption 1, we set $\gamma=0$ and replace $0\leq\xi<1/2-\gamma$ with $\forall c\in[0,1/2)$ in Lemma 14 of Andrews (2002). Since Assumptions 1 and 3 of Andrews (2002) hold under our Assumptions 1 and 3, the Lemma holds by the proof of Lemma 14 of Andrews (2002). \qed

%%%%%%%%%%%%%%%%%%%%%%%%%%%%%%%%%%%%%%%%%%%%%%%%%%%%%
\begin{lemma}
For $j=1,2$,
(a) Suppose Assumptions 1-4 hold with $d_{1}\geq 2a+2$, where $2a$ is some nonnegative integer. Then,
\[\lim_{n\rightarrow\infty}n^{a}\sup_{z\in\mathbf{R}}\left|P(T_{MR(j)}\leq z)-\left[1+\sum_{i=1}^{2a}n^{-i/2}\pi_{i}(\delta,\nu_{a})\right]\Phi(z)\right|=0.\]
(b) Suppose Assumptions 1-4 hold with $d_{1}\geq 2a+2$ and $d_{2}\geq 2a+1$, where $2a$ is some nonnegative integer. Then,
\[\lim_{n\rightarrow\infty}n^{a}P\left(\sup_{z\in\mathbf{R}}\left|P^{*}(T_{MR(j)}^{*}\leq z)-\left[1+\sum_{i=1}^{2a}n^{-i/2}\pi_{i}(\delta,\nu_{n,a}^{*})\right]\Phi(z)\right|>n^{-a}\right)=0.\]
\label{L12}
\end{lemma}
\vspace{-3em}
\noindent
\textit{Proof.}
By Lemma \ref{L10} for $\Delta_{n}=T_{MR(j)}$ and $\Delta_{n}^{*}=T_{MR(j)}^{*}$, it suffices to show that $n^{1/2}A(S_{n})$ and $n^{1/2}A(S_{n}^{*})$ possess Edgeworth expansions with remainder $o(n^{-a})$, where $A(\cdot)$ is an infinitely differentiable real-valued function. The function $A(\cdot)$ is normalized so that the asymptotic variances of $n^{1/2}A(S_{n})$ and $n^{1/2}A(S_{n}^{*})$ are one.\footnote{Hall and Horowitz (1996) and Andrews (2002) do this normalization by recentering, but the procedure is implicit.} To see this, observe that the asymptotic variances of $n^{1/2}A(S_{n})$ and $T_{MR(j)}$ are the same by Lemma \ref{L10}(a), and the conditional asymptotic variances of $n^{1/2}A(S_{n}^{*})$ and $T_{MR(j)}^{*}$ are the same, except if $\chi_{n}$ is in a sequence of sets with probability $o(n^{-a})$ by Lemma \ref{L10}(b). By Theorem 1 and 2 of Hall and Inoue (2003), the asymptotic variance of $T_{MR(j)}$ is one for $j=1,2$. To find the conditional asymptotic variance of $T_{MR(j)}^{*}$, we use the proof of Theorem 2.1. of Bickel and Freedman (1981). Conditional on $\chi_{n}$, where $\chi_{n}$ is in a sequence of sets with $P$ probability $1-o(n^{-a})$,  the usual central limit theorem and the law of large numbers imply
\begin{equation}
\sqrt{n}(\hat{\theta}_{(j)}^{*}-\hat{\theta}_{(j)})\rightarrow_{d}N(0,\Sigma_{MR(j)|F_{n}}),
\end{equation}
and $\hat{\Sigma}_{MR(j)}^{*}\rightarrow_{p}\Sigma_{MR(j)|F_{n}}$ as the resample size grows, where $\Sigma_{MR(j)|F_{n}}$ is obtained by replacing the population moments by the sample moments in the formula of $\Sigma_{MR(j)}$. By Slutsky's theorem, $T_{MR(j)}^{*}$ has the asymptotic variance of one for $j=1,2$, conditional on $\chi_{n}$, where $\chi_{n}$ is in a sequence of sets with $P$ probability $1-o(n^{-a})$. The rest of the proof is analogous to that of Lemma 16 of Andrews (2002) except that we use $n^{1/2}A(\cdot)$ in place of his $N^{1/2}G(\cdot)$. \qed

%%%%%%%%%%%%%%%%%%%%%%%%%%%%%%%%%%%%%%%%%%%%%%%%%%%%%%%%
\section*{References}
%%%%%%%%%%%%%%%%%%%%%%%%%%%%%%%%%%%%%%%%%%%%%%%%%%%%%%%%
\begin{description}
\setlength{\itemsep}{0em}
\setlength{\parskip}{0em}
\setlength{\parsep}{0em}
  \item[] Ag\"{u}ero, J. M., Marks, M. S., 2008. Motherhood and female labor force participation: evidence from infertility shocks. American Economic Review 98, 500-504.
  \item[] Andrews, D. W. K., 2002. Higher-order improvements of a computationally attractive k-step bootstrap for extremum estimators. Econometrica 70 (1), 119-162.
  \item[] Beran, R., 1988. Prepivoting test statistics: a bootstrap view of asymptotic refinements. Journal of the American Statistical Association 83 (403), 687-697.
  \item[] Berkowitz, D., Caner, M., Fang, Y., 2008. Are nearly exogenous instruments reliable? Economics Letters 101 (1), 20-23.
  \item[] Berkowitz, D., Caner, M., Fang, Y., 2012. The validity of instruments revisited. Journal of Econometrics 166 (2), 255-266.
  \item[] Bickel, P. J., Freedman, D. A., 1981. Some asymptotic theory for the bootstrap. The Annals of Statistics 9 (6), 1196-1217.
  \item[] Blundell, R., Bond, S., 1998. Initial conditions and moment restrictions in dynamic panel data models. Journal of Econometrics 87 (1), 115-143.
  \item[] Bond, S., Windmeijer, F., 2005. Reliable inference for GMM estimators? Finite sample properties of alternative test procedures in linear panel data models. Econometric Reviews 24 (1), 1-37.
  \item[] Bonnet, C., Dubois, P., 2010. Inference on vertical contracts between manufacturers and retailers allowing for nonlinear pricing and resale price maintenance. The RAND Journal of Economics 41 (1), 139-164.
  \item[] Bravo, F., 2010. Efficient M-estimators with auxiliary information. Journal of Statistical Planning and Inference 140 (11), 3326-3342.
  \item[] Brown, B. W., Newey, W. K., 2002. Generalized method of moments, efficient bootstrapping, and improved inference. Journal of Business \& Economic Statistics 20 (4), 507-517.
  \item[] Chen, X., Hong, H., Shum, M., 2007. Nonparametric likelihood ratio model selection tests between parametric likelihood and moment condition models. Journal of Econometrics 141 (1), 109-140.
  \item[] Corradi, V., Swanson, N. R., 2006. Bootstrap conditional distribution tests in the presence of dynamic misspecification. Journal of Econometrics 133 (2), 779-806.
  \item[] DiTraglia, F. J., 2012. Using invalid instruments on purpose: Focused moment selection and averaging for GMM. Working Paper. University of Pennsylvania.
  \item[] French, E., Jones, J. B., 2004. On the distribution and dynamics of health care costs. Journal of Applied Econometrics 19 (6), 705-721.
  \item[] Gallant, A. R., White, H., 1988. A unified theory of estimation and inference for nonlinear dynamic models. New York: Basil Blackwell.
  \item[] Gon{\c{c}}alves, S., White, H., 2004. Maximum likelihood and the bootstrap for nonlinear dynamic models.
Journal of Econometrics 119 (1), 199-219.
  \item[] Gowrisankaran, G., Rysman, M., 2009. Dynamics of consumer demand for new durable goods. No. w14737. National Bureau of Economic Research.
  \item[] Guggenberger, P., 2008. Finite sample evidence suggesting a heavy tail problem of the generalized empirical likelihood estimator. Econometric Reviews 27 (4-6), 526-541.
  \item[] Guggenberger, P., 2012. On the asymptotic size distortion of tests when instruments locally violate the exogeneity assumption. Econometric Theory 28 (2), 387-421.
  \item[] Guggenberger, P., Kumar, G., 2012. On the size distortion of tests after an overidentifying restrictions pretest. Journal of Applied Econometrics 27 (7), 1138-1160.
  \item[] Hahn, J., 1996. A note on bootstrapping generalized method of moments estimators. Econometric Theory 12, 187-197.
  \item[] Hahn, J., Hausman, J., 2002. A new specification test for the validity of instrumental variables. Econometrica 70 (1), 163-189.
  \item[] Hahn, J., Hausman, J., 2005. Estimation with valid and invalid instruments. Annales d'Economie et de Statistique 79-80, 25-57.
  \item[] Hall, A. R., 2000. Covariance matrix estimation and the power of the overidentifying restrictions test. Econometrica 68 (6), 1517-1527.
  \item[] Hall, A. R., 2005. Generalized method of moments. Oxford: Oxford University Press.
  \item[] Hall, A. R., Inoue, A., 2003. The large sample behavior of the generalized method of moments estimator in misspecified models. Journal of Econometrics 114 (2), 361-394.
  \item[] Hall, A. R., Pelletier, D., 2011. Non-nested testing in models estimated via generalized method of moments. Econometric Theory 27, 443-456.
  \item[] Hall, P., 1988. On symmetric bootstrap confidence intervals. Journal of the Royal Statistical Society. Series B (Methodological), 35-45.
  \item[] Hall, P., 1992. The bootstrap and Edgeworth expansion. New York: Springer-Verlag.
  \item[] Hall, P., Horowitz, J. L., 1996. Bootstrap critical values for tests based on generalized-method-of-moments estimators. Econometrica 64, 891-916.
  \item[] Hansen, L. P., 1982. Large sample properties of generalized method of moments estimators. Econometrica 50, 1029-1054.
  \item[] Hansen, L. P., Heaton, J., Yaron, A., 1996. Finite-sample properties of some alternative GMM estimators. Journal of Business \& Economic Statistics 14 (3), 262-280.
  \item[] Hellerstein, J. K., Imbens, G. W., 1999. Imposing moment restrictions from auxiliary data by weighting. Review of Economics and Statistics 81 (1), 1-14.
  \item[] Horowitz, J. L., 2001. The bootstrap. Handbook of Econometrics, Vol. 5, 3159-3228.
  \item[] Imbens, G. W., 1997. One-step estimators for over-identified generalized method of moments models. The Review of Economic Studies 64 (3), 359-383.
  \item[] Imbens, G. W., 2002. Generalized method of moments and empirical likelihood. Journal of Business \& Economic Statistics 20 (4), 493-506.
  \item[] Imbens, G. W., Lancaster, T., 1994. Combining micro and macro data in microeconometric models. The Review of Economic Studies 61 (4), 655-680.
  \item[] Inoue, A., Shintani, M., 2006. Bootstrapping GMM estimators for time series. Journal of Econometrics 133 (2), 531-555.
  \item[] Jondeau, E., Le Bihan, H., Galles, C., 2004. Assessing generalized method-of-moments estimates of the federal reserve reaction function. Journal of Business \& Economic Statistics 22 (2), 225-239.
  \item[] Kitamura, Y., 2003. A likelihood-based approach to the analysis of a class of nested and non-nested models. Working Paper. University of Pennsylvania.
  \item[] Kline, P., Santos, A., 2012. Higher order properties of the wild bootstrap under misspecification. Journal of Econometrics 171 (1), 54-70.
  \item[] Kocherlakota, N. R., 1990. On tests of representative consumer asset pricing models. Journal of Monetary Economics 26 (2), 285-304.
  \item[] Maasoumi, E., Phillips, P. C. B., 1982. On the behavior of inconsistent instrumental variable estimators. Journal of Econometrics 19 (2), 183-201.
  \item[] Marmer, V., Otsu, T., 2012. Optimal comparison of misspecified moment restriction models under chosen measure of fit. Journal of Econometrics 170, 538-550.
  \item[] Otsu, T., 2011. Moderate deviations of generalized method of moments and empirical likelihood estimators. Journal of Multivariate Analysis 102 (8), 1203-1216.
  \item[] Owen, A. B., 2001. Empirical Likelihood. Chapman and Hall/CRC.
  \item[] Parker, J. A., Julliard, C., 2005. Consumption risk and the cross section of expected returns. Journal of Political Economy 113 (1), 185-222.
  \item[] Rivers, D., Vuong, Q., 2002. Model selection tests for nonlinear dynamic models. The Econometrics Journal 5 (1), 1-39.
  \item[] Sawa, T., 1978. Information criteria for discriminating among alternative regression models. Econometrica 46, 1273-1291.
  \item[] Schennach, S. M., 2007. Point estimation with exponentially tilted empirical likelihood. The Annals of Statistics 35 (2), 634-672.
  \item[] Shi, X., 2013. A nondegenerate Vuong test. Working Paper. University of Wisconsin-Madison.
  \item[] Tauchen, G., 1986. Statistical properties of generalized method-of-moments estimators of structural parameters obtained from financial market data. Journal of Business \& Economic Statistics 4 (4), 397-416.
  \item[] Vuong, Q., 1989. Likelihood ratio tests for model selection and non-nested hypotheses. Econometrica 57, 307-333.
  \item[] White, H., 1982. Maximum likelihood estimation of misspecified models. Econometrica 50, 1-25.
  \item[] White, H., 1996. Estimation, inference and specification analysis. Vol. 22, Cambridge University Press.
\end{description}

\newpage
\thispagestyle{empty}

\begin{table}[p!]
\centering
\begin{tabular}{cccccc}
\toprule
  & Intercept      & $Edu$           & $Age-35$          & $(Age-35)^{2}$     &  $J$ test\\
  & $\theta_{0}$ & $\theta_{1}$ & $\theta_{2}$ & $\theta_{3}$ & $\chi^{2}(5)$\\
\midrule
ML &$\underset{(.317)}{1.44^{*}}$ & $\underset{(.093)}{-.009}$ & $\underset{(.015)}{-.002}$ & $\underset{(.002)}{-.002}$ & -\\
GMM &$\underset{(.268)}{1.86^{*}}$ & $\underset{(.084)}{-.109}$ & $\underset{(.002)}{-.003}$ & $\underset{(.0003)}{-.003^{*}}$ & $\underset{[.044]}{11.4}$\\
\bottomrule
\multicolumn{6}{l}{\footnotesize{Note: Standard errors in parentheses. $p$-value in bracket.}} \\
\multicolumn{6}{l}{\footnotesize{$*$: significant at 1\% level}}\\
\end{tabular}
\caption{Tables II and V of Imbens and Lancaster (1994)}
\label{IL1994}
\end{table}

\clearpage
\thispagestyle{empty}
\begin{table}[p!]
\centering
\begin{tabular}{ccccc}
\toprule
        & \multicolumn{2}{c}{Correct Model}  & \multicolumn{2}{c}{Misspecified Model}\\
\cmidrule(r){2-5}
\multirow{2}{*}{$t$ test/CI$\dag$}       & First-order   & Asymptotic        & First-order       & Asymptotic\\
              & Validity      & Refinements       & Validity          & Refinements\\
\toprule
MR         & \multirow{2}{*}{Y} & \multirow{2}{*}{Y}  & \multirow{2}{*}{Y}  & \multirow{2}{*}{Y} \\
Bootstrap$\ddag$\\
\midrule
Hall-Inoue  & \multirow{2}{*}{Y} & \multirow{2}{*}{-}  & \multirow{2}{*}{Y}  & \multirow{2}{*}{-} \\
Asymptotic\\
\midrule
\midrule
Conventional  & \multirow{2}{*}{Y} & \multirow{2}{*}{-}  & \multirow{2}{*}{-}  & \multirow{2}{*}{-} \\
Asymptotic\\
\midrule
Naive     & \multirow{2}{*}{Y} & \multirow{2}{*}{-}  & \multirow{2}{*}{-}  & \multirow{2}{*}{-} \\
Bootstrap\\
\midrule
Recentered  & \multirow{2}{*}{Y} & \multirow{2}{*}{Y}  & \multirow{2}{*}{-}  & \multirow{2}{*}{-} \\
Bootstrap\\
\bottomrule
\multicolumn{5}{l}{\footnotesize{$\dag$: The critical values are for symmetric two-sided $t$ tests and confidence intervals.}}\\
\multicolumn{5}{l}{\footnotesize{$\ddag$: MR bootstrap denotes the misspecification-robust bootstrap proposed by the author.}}
\end{tabular}
\caption{Comparison of the Asymptotic and Bootstrap Critical Values}
\label{COMP}
\end{table}

\clearpage

\thispagestyle{empty}

\begin{table}[p!]
\centering
\begin{tabular}{cccccccc}
\toprule
Degree of &          & \multicolumn{2}{c}{$n=50$}  & \multicolumn{2}{c}{$n=200$} & \multicolumn{2}{c}{$n=1000$}\\
\cmidrule(r){2-8}
\cmidrule(r){2-8}
Misspecification & Nominal Value                 & 0.90 & 0.95                 & 0.90 & 0.95 & 0.90 & 0.95\\
\toprule
                         &$CI_{MR}^{*}$        & 0.799& 0.863               & 0.848 & 0.896   & 0.887 & 0.933\\
                         &$CI_{MR}$            & 0.743 & 0.787               & 0.824 & 0.868   & 0.872 & 0.923\\
\cmidrule(r){2-8}
$\delta=0$               &$CI_{C}$             & 0.740 & 0.789               & 0.823 & 0.868   & 0.871 & 0.923\\
(correct                 &$CI_{HH}^{*}$        & 0.807 & 0.865               & 0.851 & 0.898   & 0.888 & 0.934\\
specification)           &$CI_{BN}^{*}$        & 0.806 & 0.862               & 0.850 & 0.898   &0.887&0.935\\
\cmidrule(r){2-8}
& $J$ ($J^{*}$) at 5\% & \multicolumn{2}{c}{4.7\% (4.7\%)} & \multicolumn{2}{c}{5.2\% (5.5\%)} & \multicolumn{2}{c}{5.3\% (5.5\%)}\\
\bottomrule
                         &$CI_{MR}^{*}$        & 0.783 & 0.842               & 0.834 & 0.893   & 0.873 & 0.919\\
                         &$CI_{MR}$            & 0.715 & 0.761               & 0.801 & 0.852   & 0.854 & 0.904\\
\cmidrule(r){2-8}
$\delta=-0.3$             &$CI_{C}$             & 0.633 & 0.692               & 0.692 & 0.764   & 0.716 & 0.797\\
(moderate                &$CI_{HH}^{*}$        & 0.728 & 0.799               & 0.757 & 0.825  & 0.755 & 0.837\\
misspecification)        &$CI_{BN}^{*}$        & 0.706 & 0.783               & 0.744 & 0.816   & 0.749 & 0.832\\
\cmidrule(r){2-8}
& $J$ ($J^{*}$) at 5\%  & \multicolumn{2}{c}{55.2\% (55.3\%)} & \multicolumn{2}{c}{99.1\% (99.0\%)} & \multicolumn{2}{c}{100\% (100\%)}\\
\bottomrule
                         &$CI_{MR}^{*}$        & 0.777 & 0.834               & 0.824 & 0.877   & 0.861 & 0.910\\
                         &$CI_{MR}$            & 0.701 & 0.753               & 0.788 & 0.836   & 0.844 & 0.892\\
\cmidrule(r){2-8}
$\delta=-0.6$             &$CI_{C}$             & 0.521 & 0.597               & 0.561 & 0.636   & 0.576 & 0.662\\
(large                &$CI_{HH}^{*}$        & 0.674 & 0.747               & 0.656 & 0.750   & 0.635 & 0.732\\
misspecification)        &$CI_{BN}^{*}$        & 0.612 & 0.709               & 0.614 & 0.716  & 0.539 & 0.628\\
\cmidrule(r){2-8}
& $J$ ($J^{*}$) at 5\%  & \multicolumn{2}{c}{98.6\% (98.4\%)} & \multicolumn{2}{c}{100\% (100\%)} & \multicolumn{2}{c}{100\% (100\%)}\\
\bottomrule
                         &$CI_{MR}^{*}$        & 0.893 & 0.936               & 0.915 & 0.957   &0.916&0.961\\
                         &$CI_{MR}$            & 0.864 & 0.914               & 0.900 & 0.949  &0.906& 0.956\\
\cmidrule(r){2-8}
$\delta=0.6$             &$CI_{C}$             & 0.925 & 0.958               & 0.972 & 0.989   &0.988&0.997\\
(large                &$CI_{HH}^{*}$        & 0.960 & 0.983               & 0.982 & 0.994   &0.991&0.998\\
misspecification)        &$CI_{BN}^{*}$        & 0.954 & 0.973               & 0.941 & 0.950  &0.685&0.689 \\
\cmidrule(r){2-8}
& $J$ ($J^{*}$) at 5\%  & \multicolumn{2}{c}{98.6\% (98.6\%)} & \multicolumn{2}{c}{100\% (100\%)} & \multicolumn{2}{c}{100\% (100\%)}\\
\bottomrule
\end{tabular}
\caption{Coverage Probabilities of 90\% and 95\% Confidence Intervals for $\theta_{0(2)}$ based on the Two-step GMM Estimator, $\hat{\theta}_{(2)}$, in Example 1 DGP \eqref{DGP1}. $r= 5,000$ and $B=1,000$. $J$ and $J^{*}$ at 5\% denote the rejection probabilities of the asymptotic and the HH bootstrap $J$ test at 5\% level, respectively.}
\label{table_ex1}
\end{table}

\clearpage

\thispagestyle{empty}

\begin{table}[p!]
\centering
\begin{tabular}{cccccccc}
\toprule
Degree of &          & \multicolumn{2}{c}{$n=50$}  & \multicolumn{2}{c}{$n=200$} & \multicolumn{2}{c}{$n=1000$}\\
\cmidrule(r){2-8}
\cmidrule(r){2-8}
Misspecification & Nominal Value                 & 0.90 & 0.95                 & 0.90 & 0.95 & 0.90 & 0.95\\
\toprule
                         &$CI_{MR}^{*}$        & 0.647 & 0.726               & 0.786 & 0.852   & 0.862 & 0.914\\
                         &$CI_{MR}$            & 0.526 & 0.578               & 0.728 & 0.781   & 0.851 & 0.891\\
\cmidrule(r){2-8}
$\delta=0$               &$CI_{C}$             & 0.425 & 0.473               & 0.647 & 0.701   & 0.838 & 0.880\\
(correct                 &$CI_{HH}^{*}$        & 0.584 & 0.650               & 0.750 & 0.809   & 0.859 & 0.912\\
specification)           &$CI_{BN}^{*}$        & 0.576 & 0.648               & 0.727 & 0.790   & 0.858 & 0.911\\
\cmidrule(r){2-8}
& $J$ ($J^{*}$) at 5\% & \multicolumn{2}{c}{27.4\% (17.4\%)} & \multicolumn{2}{c}{20.8\% (12.0\%)} & \multicolumn{2}{c}{8.8\% (4.9\%)}\\
\bottomrule
                         &$CI_{MR}^{*}$        & 0.759 & 0.827               & 0.850 & 0.901   & 0.937 & 0.961\\
                         &$CI_{MR}$            & 0.653 & 0.703               & 0.807 & 0.845   & 0.909 & 0.937\\
\cmidrule(r){2-8}
$\delta=0.25$            &$CI_{C}$             & 0.535 & 0.586               & 0.663 & 0.694   & 0.736 & 0.769\\
(small                   &$CI_{HH}^{*}$        & 0.675 & 0.737               & 0.742 & 0.785   & 0.778 & 0.822\\
misspecification)        &$CI_{BN}^{*}$        & 0.666 & 0.736               & 0.706 & 0.752   & 0.746 & 0.787\\
\cmidrule(r){2-8}
& $J$ ($J^{*}$) at 5\% & \multicolumn{2}{c}{33.5\% (21.4\%)} & \multicolumn{2}{c}{43.6\% (28.4\%)} & \multicolumn{2}{c}{77.7\% (58.6\%)}\\
\bottomrule
                         &$CI_{MR}^{*}$        & 0.866 & 0.904               & 0.904 & 0.934   & 0.896 & 0.958\\
                         &$CI_{MR}$            & 0.778 & 0.815               & 0.839 & 0.868   & 0.810 & 0.893\\
\cmidrule(r){2-8}
$\delta=0.5$             &$CI_{C}$            & 0.672 & 0.711                & 0.687 & 0.713   & 0.539 & 0.659\\
(moderate                &$CI_{HH}^{*}$        & 0.774 & 0.816               & 0.783 & 0.815   & 0.715 & 0.769\\
misspecification)        &$CI_{BN}^{*}$        & 0.770 & 0.821               & 0.738 & 0.778   & 0.649 & 0.711\\
\cmidrule(r){2-8}
& $J$ ($J^{*}$) at 5\% & \multicolumn{2}{c}{33.8\% (20.2\%)} & \multicolumn{2}{c}{50.7\% (29.3\%)} & \multicolumn{2}{c}{90.6\% (64.9\%)}\\
\bottomrule
\end{tabular}
\caption{Coverage Probabilities of 90\% and 95\% Confidence Intervals for $\beta_{0(1)}$ based on the One-step GMM Estimator, $\hat{\beta}_{(1)}$ in Example 2 DGP \eqref{DGP2}. $r= 5,000$ and $B=1,000$. $J$ and $J^{*}$ at 5\% denote the rejection probabilities of the asymptotic and the HH bootstrap $J$ test at 5\% level, respectively.}
\label{table_ex2}
\end{table}

\clearpage
\thispagestyle{empty}

\clearpage 
\thispagestyle{empty}

\clearpage
\thispagestyle{empty}

\clearpage
\thispagestyle{empty}

\end{document}